\documentclass[12pt,manuscript]{aastex}
\usepackage{apjfonts}

\newcommand{\msun}{M_\odot}
\newcommand{\rsun}{R_\odot}

\newcommand{\gapprox}{\mathrel{\mathpalette\@versim>}}
\newcommand{\lapprox}{\mathrel{\mathpalette\@versim<}}
\newcommand{\propapprox}{\mathrel{\mathpalette\@versim\propto}}

\shorttitle{RCW 86: A Type Ia SN in a Bubble}
\shortauthors{WILLIAMS ET AL.}

\begin{document}

\title{RCW 86: A Type Ia Supernova in a Wind-Blown Bubble}

\author{Brian J. Williams,\altaffilmark{1}
William P. Blair,\altaffilmark{2}
John M. Blondin,\altaffilmark{1}
Kazimierz J. Borkowski,\altaffilmark{1}
Parviz Ghavamian,\altaffilmark{3}
Knox S. Long,\altaffilmark{3}
John C. Raymond,\altaffilmark{4}
Stephen P. Reynolds,\altaffilmark{1}
Jeonghee Rho,\altaffilmark{5}
\& P. Frank Winkler\altaffilmark{6}
}

\altaffiltext{1}{Physics Dept., North Carolina State University,
    Raleigh, NC 27695-8202; bjwilli2@ncsu.edu.}
\altaffiltext{2}{Dept. of Physics and Astronomy, Johns Hopkins University, 
    3400 N. Charles St., Baltimore, MD 21218-2686.}
\altaffiltext{3}{STScI, 3700 San Martin Dr., Baltimore, MD 21218.}
\altaffiltext{4}{Harvard-Smithsonian Center for Astrophysics, 60 Garden
    Street, Cambridge, MA 02138.}
\altaffiltext{5}{SOFIA/USRA}
\altaffiltext{6}{Dept. of Physics, Middlebury College, Middlebury, VT 
    05753.}

\begin{abstract}

We report results from a multi-wavelength analysis of the Galactic SNR
RCW 86, the proposed remnant of the supernova of 185 A.D. We report
new infrared observations from {\it Spitzer} and {\it WISE}, where the
entire shell is detected at 24 and 22 $\mu$m. We fit the infrared flux
ratios with models of collisionally heated ambient dust, finding
post-shock gas densities in the non-radiative shocks of 2.4 and 2.0
cm$^{-3}$ in the SW and NW portions of the remnant, respectively. The
Balmer-dominated shocks around the periphery of the shell, large
amount of iron in the X-ray emitting ejecta, and lack of a compact
remnant support a Type Ia origin for this remnant. From hydrodynamic
simulations, the observed characteristics of RCW 86 are successfully
reproduced by an off-center explosion in a low-density cavity carved
by the progenitor system.  This would make RCW 86 the first known case
of a Type Ia supernova in a wind-blown bubble. The fast shocks ($>
3000$ km s$^{-1}$) observed in the NE are propagating in the
low-density bubble, where the shock is just beginning to encounter the
shell, while the slower shocks elsewhere have already encountered the
bubble wall. The diffuse nature of the synchrotron emission in the SW
and NW is due to electrons that were accelerated early in the lifetime
of the remnant, when the shock was still in the bubble. Electrons in a
bubble could produce gamma-rays by inverse-Compton scattering. The
wind-blown bubble scenario requires a single-degenerate progenitor,
which should leave behind a companion star.

\keywords{
interstellar medium: dust ---
supernova remnants ---
}

\end{abstract}

\section{Introduction}
\label{intro}

It is widely believed that Type Ia supernovae (SNe) originate from
white dwarf (WD) stars that have been pushed close to the
Chandrasekhar limit, either by accreting matter from a companion star
(single-degenerate scenario), or by merging with another WD
(double-degenerate scenario) \citep{isern08}. The detailed physics of
the single-degenerate scenario, particularly the role, if any, that
the binary system plays in shaping the surrounding medium is poorly
understood. \citet{hachisu96} suggested that, under certain
conditions, accreting WDs in a binary system with a main sequence or
red giant companion might blow a substantial wind, carving out
low-density cavities in the surrounding ISM in the $\sim 10^{6}$ yr
prior to explosion. \citet{badenes07} considered young ($\le 1000$ yr)
supernova remnants (SNRs) known to be of Type Ia origin, finding that
none were consistent with evolution into such a modified medium.

The Galactic SNR RCW 86 (G315.4-2.3; MSH 14-63) is one of a small
number of remnants with a historical connection, as it has been
proposed to be the remains of SN 185 A.D. \citep{clark75}. The
distance has been reported to be near 1 kpc \citep{kaastra92}, but
more recent measurements have put it from 2.3 \citep{sollerman03} to
2.8 kpc \citep{rosado96}. Throughout this work, we adopt a distance of
2.5 kpc, parameterized, where appropriate, as $d_{2.5}$. The claim
that this event was a SN at all has been disputed \citep{chin94}, but
\citet{stephenson02} review the entirety of the available information
on the ``guest star'' of 185 A.D., concluding that it was indeed a SN,
a view supported by the work of \citet{zhou06}, and that RCW 86
represents the most likely candidate for its remnant. According to
\citet{schaefer93}, the maximum unrefracted altitude of RCW 86 as seen
from the city of Luoyang, China, was 2.3$^{\circ}$, leading to an
airmass of 18. Under good seeing conditions (0.2 magnitudes extinction
per airmass), the atmospheric extinction would be $A_{V}$ =
3.6. Optical measurements have placed the ISM extinction at $\sim
A_{V} = 1.7$ \citep{leibowitz83}, so a supernova at 2.5 kpc (see
below) with, for instance, $M_{V} = -17$ would appear as a new star of
$m = 0.3$, easily visible to ancient observers. Significant sources of
error are possible in virtually all of these parameters. If the
Chinese historical observations came from a region farther south in
China, the airmass in the direction of RCW 86 would be significantly
reduced. For instance, an altitude of 5$^{\circ}$ above the horizon
implies 10 airmasses \citep{kasten89}, reducing the atmospheric
extinction to 2 mag. A brighter Type Ia SN, which we argue in favor of
later, could add another 2 mag to the observed brightness. On the
other hand, we report X-ray column densities of $\sim 6 \times
10^{21}$ cm$^{-2}$, which could raise the interstellar extinction from
2 to 4 mag. In summary, it appears quite possible that there could
have been a supernova visible from China at the position of RCW 86, at
a distance of 2.5 kpc.

The recent measurement of high proper motion of the shock, as defined
by the onset of X-ray emission in the NE limb implies very fast shocks
($\sim 6000$ km s$^{-1}$) \citep{helder09}, supporting a young age for
the remnant. The broad Fe K$\alpha$ line width measured by {\it
  Suzaku} \citep{ueno07} and the presence of non-thermal synchrotron
X-ray emission \citep{bamba00,borkowski01a,vink06} also suggest a young
SNR. The remnant shows a complete shell at radio, optical, and X-ray
energies \citep{kesteven87,smith97,pisarski84}, and has recently been
detected in TeV gamma-rays by H.E.S.S. \citep{aharonian09}. Shock
speeds vary greatly from one side of the remnant to the other, from
the high shock speeds in the east to the slower (550 -- 650 km
s$^{-1}$) shocks in the NW and SW corners, as derived from optical
spectroscopy \citep{long90,ghavamian99,ghavamian01}. \citet{helder09}
also report that the eastern limb could be a site of significant
cosmic-ray acceleration, based on the discrepancy between proper
motions of the X-ray emitting filaments and the spectroscopically
determined proton temperature.

Numerous authors have pointed out the difficulties with reconciling
the young age of RCW 86 with its large size. At a distance of 2.5 kpc,
the average speed of the shock for this large remnant ($40'$ in
diameter) is $\sim 7800$ km s$^{-1}$, but as previously mentioned,
shock speeds in most of the remnant are an order of magnitude less
than this. It has been suggested (e.g., Vink et al. 1997) that a
cavity explosion offers the most natural explanation, with the shock
racing through a low-density bubble, recently having encountered a
dense shell. Cavity explosions are typically associated with
core-collapse (CC) supernovae, but the nature of the progenitor system
for RCW 86 is uncertain. \citet{badenes07} suggested that RCW 86 may
be the result of a Type Ia explosion into a cavity formed by the
progenitor system, a scenario that could offer new constraints on
accretion wind outflows from single-degenerate progenitor systems of
Type Ia SNe.

This paper is organized as follows: In Sections 2 and 3, we report the
results of new infrared (IR) observations with {\it Spitzer} and {\it
  WISE}. We discuss the nature of the progenitor system in
Section~\ref{progenitor}, and the evidence for a cavity explosion in
Section~\ref{hydro}. The nonthermal synchrotron emission is discussed
in Section 6, and the gamma-ray emission in Section 7.

\section{Infrared Observations}
\label{observations}

IR emission from SNRs can be thermal continuum from warm dust grains
heated by collisions with energetic ions and electrons in the X-ray
emitting gas, line emission from low to medium-ionization states of
atoms cooling in the post-shock region, or a combination of the
two. Nonthermal emission via synchrotron radiation can contribute as
well, but is generally negligible compared to thermal processes
\citep{rho08}. Generally speaking, emission from younger SNRs
expanding into a low-density medium (characterized by non-radiative
shocks in the optical) will be dominated by continuum
\citep{williams11}, while older shocks or those encountering molecular
clouds will become radiative and show strong mid-IR line emission
\citep{hewitt09}. IR studies of SNRs offer the possibility to learn
about both the conditions of the remnant and the physics of the
gas-dust interaction in shocks.

Previous IR observations of RCW 86 have led to conflicting
results. \citet{dwek87b} used IRAS observations to determine that the
IR/X-ray flux ratio (IRX ratio) from the bright SW shell (the only
part of the remnant firmly detected by IRAS) is higher than
theoretical predictions. They attributed this to either a dust/gas
mass ratio higher than average for the Galaxy or a poor coupling
between the IR emitting dust and the X-ray emitting gas. However,
\citet{greidanus90} used the same IRAS data and IR emission model to
arrive at the opposite conclusion, namely that the IRX ratio for the
remnant is lower than predictions by a factor of 10, implying a low
dust/gas mass ratio.

Studies done with IRAS suffer from both poor angular resolution and
low sensitivity, and resolving discrepant results such as these
requires study by advanced IR telescopes. Using {\it Spitzer} and {\it
  WISE}, we detect the entire shell at 24 and 22 $\mu$m and the SW and
NW portions at 12 and 70 $\mu$m. Faint extended emission in the SW is
seen in short-wavelength bands as well. Detection of the eastern shell
of the remnant is significantly complicated at 70 $\mu$m by confusion
with Galactic IR emission.

As part of {\it Spitzer} program 50698, we obtained a full map of the
remnant at 24 and 70 $\mu$m using the Multiband Imaging Photometer for
Spitzer (MIPS) on 2009 March 31, just weeks before the end of the cold
{\it Spitzer} mission. The size of the remnant ($\sim 40'$ in
diameter) required two scan maps (AOR ID 26750208 \& 26749952) of
1$^\circ$ $\times$ 0.5$^\circ$, which we mosaicked together using the
MOPEX software provided by the Spitzer Science Center. We used a
medium scan rate for the maps, offsetting each scan leg by $148''$ to
ensure full coverage at 70 $\mu$m and redundancy by a factor of 2 at
24 $\mu$m. We used all four channels (from 3.6 to 8 $\mu$m) of the
Infrared Array Camera (IRAC) to map a region of $\sim 7' \times 7'$ on
2009 March 23 (AOR ID 26748928), centered on the brightest filaments
in the SW region of the shell seen in X-rays with {\it Chandra} (Rho
et al. 2002, hereafter R02). Both the IRAC and MIPS data were
processed using version 18.7 of the {\it Spitzer} pipeline. The
pipeline processed 70 $\mu$m data contains some artifacts, most
notably striping in the direction of the scan leg.

In mid-April of 2011, the {\it Wide-Field Infrared Survey Explorer}
(WISE) science team released the first half of the data taken by WISE,
a mid-IR telescope which surveyed the entire sky at 3.3, 4.6, 12, and
22 $\mu$m. RCW 86 was in the region of the sky covered by this initial
data release. To provide a more complete picture of the IR properties
of the SNR, we have downloaded the {\it WISE} images of this remnant
at all four wavelengths.

\section{Analysis and Modeling}
\label{analysis}

\subsection{Morphology}

We show the MIPS images at 24 and 70 $\mu$m in Figure~\ref{images},
along with X-ray and optical images as described in the caption. At 24
$\mu$m, where the angular resolution of the telescope is $\sim 7''$,
the correlation between the IR, X-ray, and optical images is
significant, implying a tight coupling between the IR emitting dust
and the X-ray emitting gas. A strong correlation exists for the 70
$\mu$m data (resolution $\sim 20''$) as well in the SW and NW portions
of the shell; however, the eastern part of the shell is dominated by
emission that is likely from a foreground or background region. The IR
colors in the E region allow us to trace the forward shock and make a
determination of what is and is not associated with the remnant. At 24
$\mu$m, the much lower background level allows distinction between
this region and the thin, non-radiative filaments of the blast wave.

We show a three-color IRAC image in Figure~\ref{4panel_irac}, where a
few faint filaments are observed that correspond to features seen at
24 $\mu$m. These filaments also appear as radiative filaments in the
[S II] image \citep{smith97}, making it likely that the IRAC emission
comes from lines (most likely [Ar II] at 7 $\mu$m and [Fe II] at 5.4
$\mu$m). It is possible that emission from polycyclic aromatic
hydrocarbons (PAHs) heated immediately behind the shock front could
contribute as well, although the lifetime of these molecules is quite
short in the post shock environment.

\begin{figure}
\includegraphics[width=15cm]{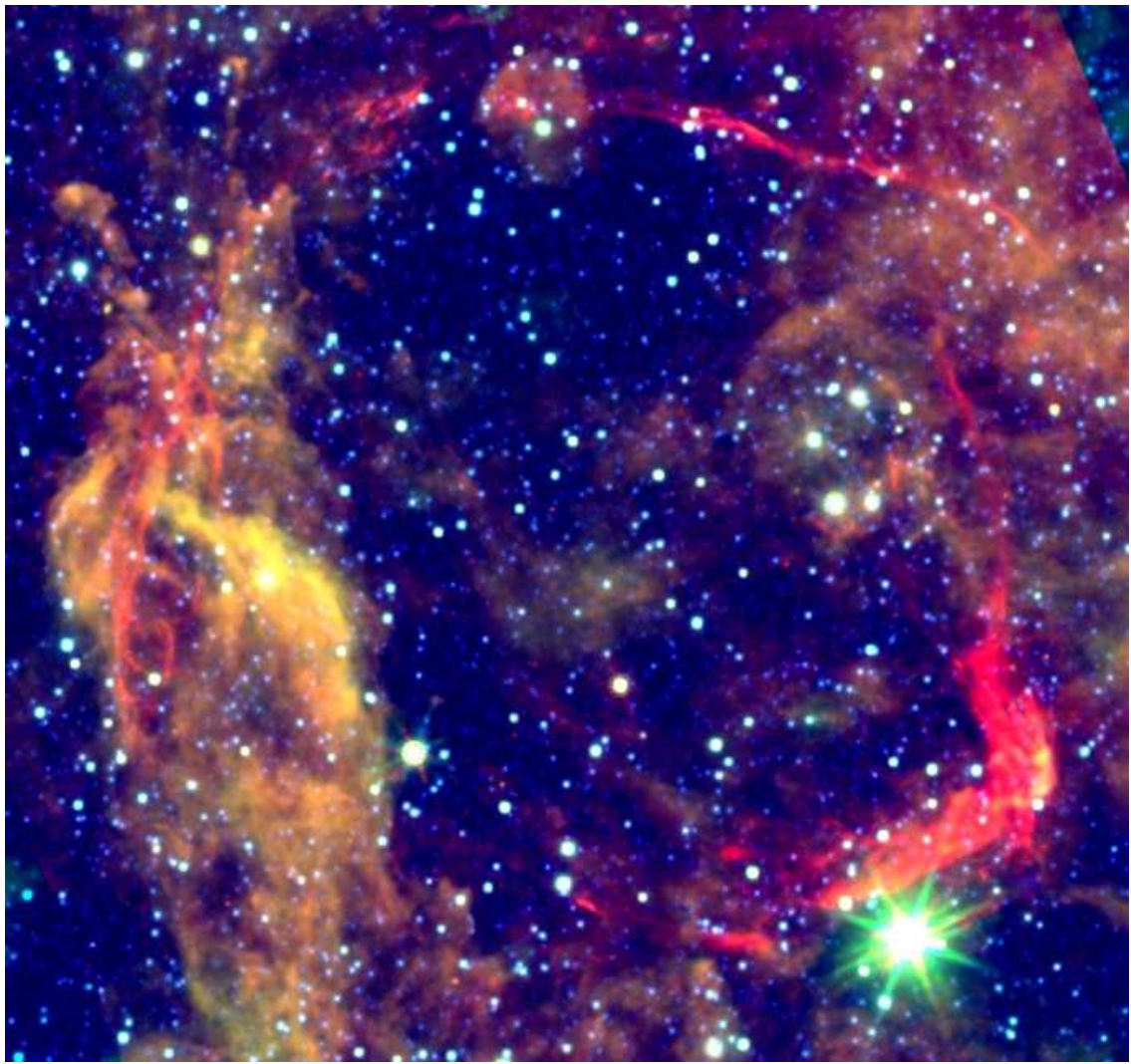}
\caption{{\it Spitzer} and {\it WISE} mosaic of the entire remnant,
  with {\it Spitzer} 24 $\mu$m emission in red, {\it WISE} 12 $\mu$m
  emission in green, and {\it WISE} 4.6 $\mu$m emission in blue. The
  22 $\mu$m image from {\it WISE}, not shown, looks virtually
  identical to the {\it Spitzer} 24 $\mu$m image. We display the MIPS
  24 $\mu$m image here because of the sharper resolution and better
  signal-to-noise ratio. Only the brightest sections of the NW and SW
  are visible at 12 $\mu$m (note the slightly yellowish colors of
  these regions).
\label{spitzerwise}
}
\end{figure}

A mosaic of the {\it WISE} 4.6 $\mu$m, 12 $\mu$m, and {\it Spitzer} 24
$\mu$m images is shown in Figure~\ref{spitzerwise}. WISE data show the
entire periphery of the shell at 22 $\mu$m (though with a lower
angular resolution than with {\it Spitzer}), as well as significant
portions of the SW shell at 12 $\mu$m, where the {\it WISE} resolution
is comparable with {\it Spitzer's} at 24 $\mu$m. Faint emission from
the NW filament is detectable at 12 $\mu$m as well. At 4.6 and 3.3
$\mu$m, only a small filament is visible in the extreme SW, a filament
which was not covered by the spatially limited IRAC map. The colors of
this image, which reflect the temperature of the dust, separate the
emission that is associated with the SNR from that of the
background. This is particularly important in the east, where the thin
red filaments trace the blast wave.

\subsection{IR Fluxes and Ratios}
\label{fluxes}

\begin{figure}
\includegraphics[width=15cm]{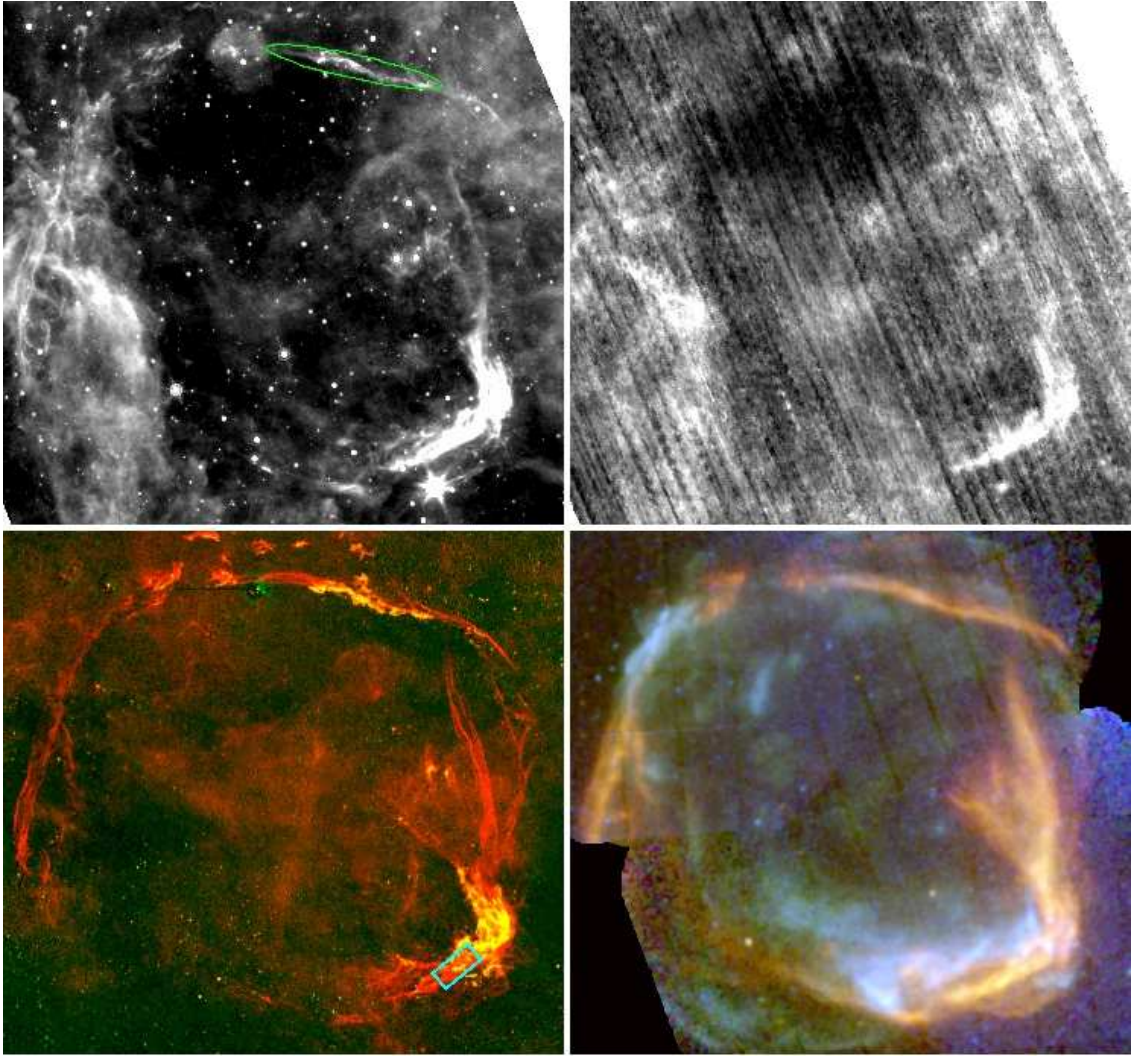}
\caption{Top Left: MIPS 24 $\mu$m image, with NW extraction region
  indicated as green ellipse; Top Right: MIPS 70 $\mu$m image; Bottom
  Right: {\it XMM-Newton} EPIC image, with 0.5-1.0 keV in red, 1 keV
  to 1.95 keV in green, and 1.95-6.8 keV in blue; Bottom Left:
  NOAO/CTIO star-subtracted optical image \citep{smith97}, with
  H$\alpha$ in red and [S II] in green. Areas appearing yellow contain
  both non-radiative and radiative shocks. Cyan rectangle (lower left
  panel) marks {\it Spitzer} IRS spectral extraction region, as
  described in text. In all images, N is up and E is to the left. The
  remnant is approximately $40'$ in diameter.
\label{images}
}
\end{figure}

\begin{figure}
\includegraphics[width=15cm]{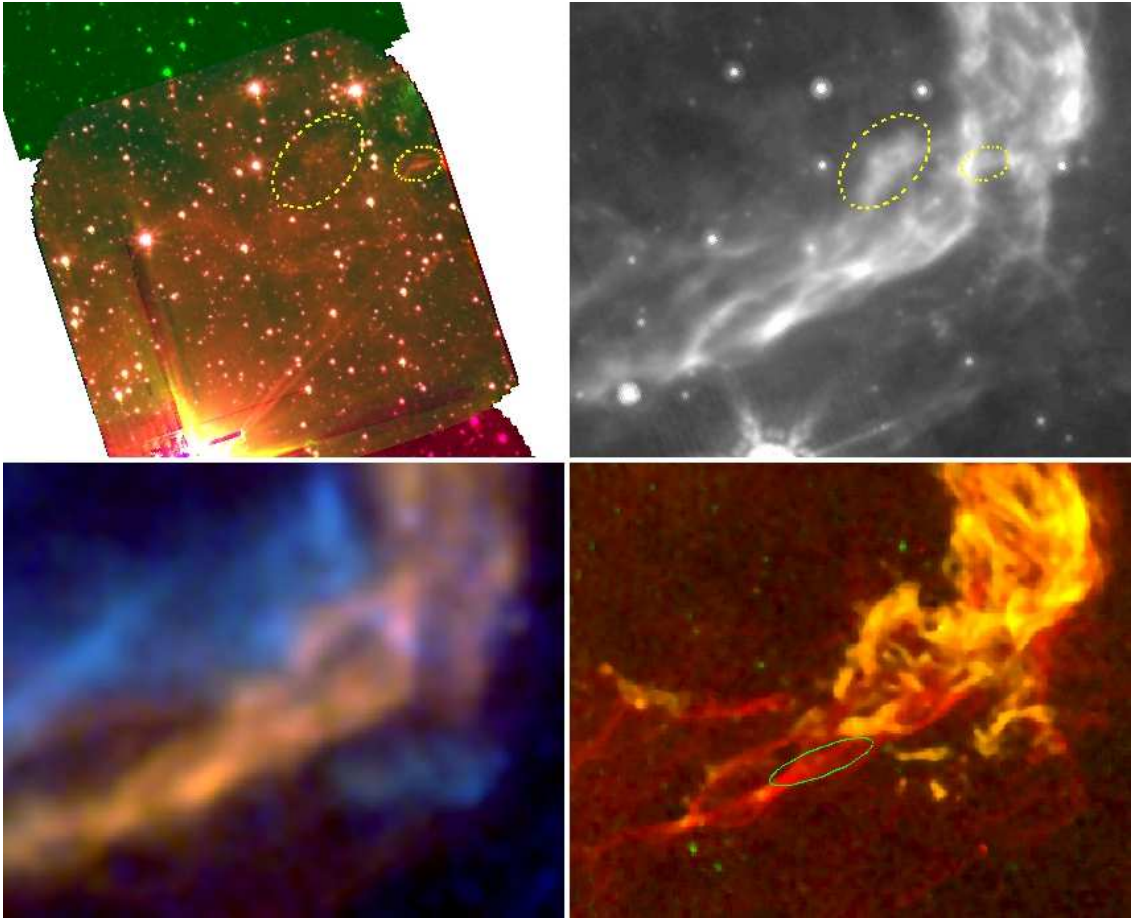}
\caption{Top Left: IRAC 3-color image of SW region, with 8 $\mu$m in
  red, 5.8 $\mu$m in green, 4.5 $\mu$m in blue. Yellow regions
  highlight structures also seen at 24 $\mu$m; Top Right: MIPS 24
  $\mu$m image with IRAC regions overlaid; Bottom Right: optical
  H$\alpha$ and [S II] image, as in Figure~\ref{images}, with region
  highlighting a purely non-radiative filament, used for IR analysis
  of the SW region, as discussed in the text; Bottom Left: EPIC-MOS2
  X-ray image, with colors as in Figure~\ref{images}.
\label{4panel_irac}
}
\end{figure}

For our analysis, we selected two regions in the remnant with
detections at both 24 and 70 $\mu$m: a small region in the bright SW
``corner'' that corresponds to purely non-radiative emission, as
identified in optical images, and a filament in the NW stretching
$\sim 7'$. The regions are indicated in Figure~\ref{images} and
Figure~\ref{4panel_irac}, and are significantly larger than the
angular resolution of {\it Spitzer} at 70 $\mu$m. To measure the flux,
we chose several local background regions outside of the remnant near
the region of interest. The background is somewhat variable,
particularly in the NW, so we average the background subtracted fluxes
and report the standard deviation of these measurements as the
error. Because we are interested in the emission from dust only, we
must account for both stars and line emission from cooling
gas. Neither region contains bright stellar sources, and only the 24
$\mu$m image shows any stars at all in the region of interest. We
excise regions at 24 $\mu$m where stars are seen, using IRAC (where
available) and optical images to determine the stellar nature of
unresolved sources. Despite no visible stars in the regions at 70
$\mu$m, we excise the same regions when measuring the 70 $\mu$m
flux. We find that stellar contamination to the 24 $\mu$m flux is
negligible in the small SW region and accounts for 15\% in the larger
NW region. The integrated flux from the excised regions at 70 $\mu$m
was less than 1\% of the total. Fluxes reported in this work are the
``star-subtracted'' fluxes. We measure fluxes in the SW of 0.37 and
1.41 Jy at 24 and 70 $\mu$m, respectively, and 0.69 and 3.10 Jy in the
NW. Errors on these flux measurements are $\sim 6\%$ in the SW and
$\sim 30\%$ in the NW, where the larger errors in the NW result from a
fainter, more diffuse source region with a more variable
background. The 70/24 $\mu$m flux ratios are 3.8$\pm 0.3$ and 4.5$\pm
2$ in the SW and NW, respectively.

To account for the contribution of line emission to the integrated
fluxes, we use IRS spectra taken as part of our {\it Spitzer}
observing program. We defer analysis of the IRS data to a future
publication, but note that a spatially integrated low-resolution
spectrum from the SW shell shows the presence of the [Fe II]/[O IV]
line blend at $\sim 26$ $\mu$m (the lines are unresolved by the
spectrograph, which has spectral resolution $\lambda/\Delta\lambda$
$\sim 100$). This line lies on top of a strong dust continuum, and
when we integrate the IRS spectrum over the MIPS 24 $\mu$m bandpass
and response curve, the total line flux accounts for 1.2\% of the
observed 24 $\mu$m flux. The extraction region for this spectrum is
shown in the bottom left panel of Figure~\ref{images}, and lies on top
of both radiative and non-radiative filaments. Although we do not have
IRS spectra of the NW filament, radiative shocks there are fainter,
and we expect that line contamination to the 24 $\mu$m fluxes should
be even less important than in the SW. We do not have spectral data
extending to the wavelengths of the 70 $\mu$m detector, but the
bandpass of that filter is extremely broad (FWHM $\sim 20\ \mu$m), and
any lines present, such as [O I] at 63 $\mu$m or [O III] at 88 $\mu$m
(where filter throughput is only 14\%) would have to be
extraordinarily strong to contribute significant flux. We thus ignore
any line contributions to the fluxes reported above.

\subsection{Modeling}
\label{modeling}

To model IR emission, we follow a procedure identical to that
described in previous works \citep{borkowski06,williams06}. The
heating rate of a grain immersed in a hot plasma depends on both the
post-shock gas density (collision rate) and temperature (energy per
collision). In general, the temperature of a plasma can be determined
by fits to X-ray spectra. Dust heating models depend on this
temperature, but this dependency is not strong \citep{dwek87}. The
density of the gas is a more sensitive parameter for dust heating
models and controls the temperature of the grains (i.e. the 70/24
$\mu$m flux ratio). We regard the gas density as a free model
parameter and adjust it to match the observed IR fluxes.

We account for sputtering of grains by ions, using a plane-shock model
that superimposes regions of increasing sputtering timescale, defined
as $\tau_{sp}=\int_0^t n_p dt$, where $n_{p}$ is the post-shock proton
density, while keeping temperature behind the shock constant. (Note
that the sputtering timescale is directly related to the ionization
timescale used in X-rays, defined as $\tau_i=\int_0^t n_e dt.$) Inputs
to the model are a pre-shock grain size distribution, grain type and
abundance (for these quantities, we use models for the Galaxy from
\citet{weingartner01}, which contain a mixture of amorphous silicate
and graphite grains), and ion and electron temperature and
density. The code calculates heating and sputtering for grains from 1
nm to 1 $\mu$m, producing a unique temperature and spectrum for each
grain size. The spectra are summed according to the post-sputtering
size distribution to produce a single model spectrum. We use
sputtering rates from \citet{nozawa06} with enhancements for small
grains from \citet{jurac98}. Although we do not include it in our
calculations, we note that the shock speeds in this remnant are only
slightly above the upper threshold, reported by \citet{sankrit10} as
500 km s$^{-1}$, where nonthermal, or ballistic sputtering effects
begin to become important.

To determine kT and $\tau_{i}$ for the X-ray emitting gas, we analyze
X-ray spectra from archival {\it Chandra} and {\it XMM-Newton}
observations. The {\it Chandra} data were obtained in 2001 (Obs. ID
1993), where the SW shell was observed for 92 ks. We reprocessed and
extracted the data using version 4.3 of the {\it Chandra} Interactive
Analysis of Observations (CIAO). We obtained archival X-ray data in
the NW rim and SW corner from the {\it XMM-Newton} EPIC-MOS detectors,
observed in 2007 (Obs. ID 0504810401 \& 0504810301, J. Vink,
P.I.). The {\it XMM-Newton} data were processed using version 9.0 of
the Science Analysis Software (SAS) for XMM. We filtered the data from
MOS 1 and MOS 2 for flares and bad pixels, grouping the data to a
minimum of 25 counts per bin. We fit the data using version 12.6.0 of
{\it XSPEC} \citep{arnaud96}. 

\subsection{SW}

R02 used the high spatial resolution of {\it Chandra} to uncover a
complex system of shocks in the SW, with hard regions from the shocked
ejecta mixed with softer regions from shocked ISM and regions
dominated by non-thermal emission. We employ a similar modeling
procedure to that used in R02 to fit the X-ray spectra from our region
in the SW. We jointly fit {\it Chandra} ACIS and {\it XMM-Newton}
EPIC-MOS 2 data (this particular region falls on the dead MOS 1 chip),
letting the abundances of C, N (tied to C), O, Mg, Si, and Fe float
freely. We find a reasonable fit with kT $= 0.39$ keV and $\tau_{i} =
6.0 \times 10^{10}$ cm$^{-3}$ s. The reduced $\chi^{2}$ value in this
simple model is rather high, $\sim 3$. R02 found similarly poor values
of $\chi^{2}$ in modeling emission from this region, attributing it to
poor Fe L-shell data in low temperature plasmas (see discussion in
Section 3.2 of that paper). Excising the regions of the spectra that
were poorly fit improved the value of $\chi^{2}$, but did not affect
the fitted parameters of the thermal model, which are fully listed in
Table~\ref{xrayparams}. Freezing these parameters ($T_{e}$ and
$\tau_{i}$) in our dust heating model, we find that the IR luminosity
and flux ratio are best fit with a post-shock density of $n_{H}$ = 2.4
(2.1, 2.75) cm$^{-3}$ and a radiating dust mass of 3.2 $d_{2.5}^2$
(2.7, 3.9) $\times 10^{-4}$ $\msun$. Lower and upper limits, listed in
parentheses, are derived from fitting dust emission models to the
extrema of the $S_{70}/S_{24}$ ratio. We assume that $T_{p} = T_{e} =
0.39$ keV, which is appropriate for the shock speeds and ages seen in
this section of the remnant \citep{ghavamian07,vanadelsberg08}. In any
case, grain heating at these temperatures is done almost entirely by
electrons \citep{dwek87}.

\subsection{NW}

\begin{figure}
\includegraphics[width=15cm]{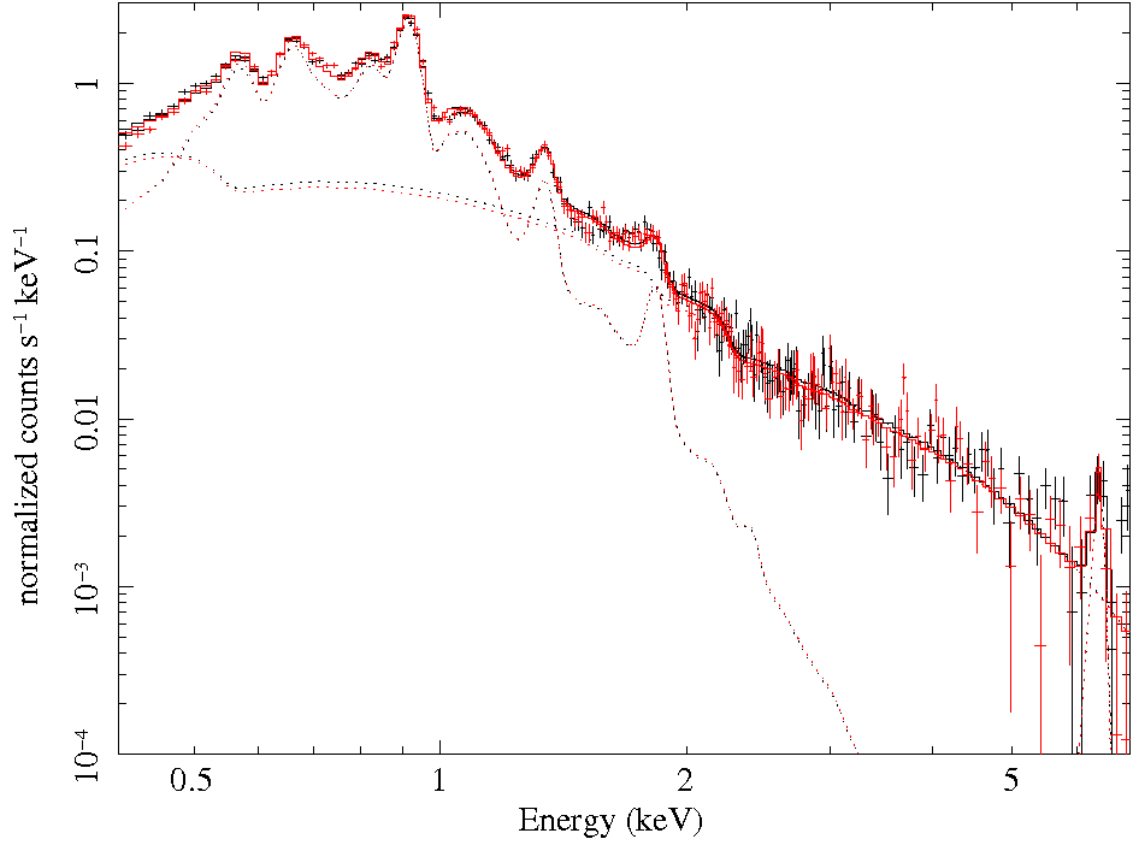}
\caption{EPIC-MOS 1 and 2 spectra, in black and red, respectively, of
  NW region, overlaid with a model of thermal and non-thermal
  emission, as described in Section~\ref{modeling}. An additional
  thermal model to account for the Fe K$\alpha$ line at 6.4 keV is
  included as well. Data are 3-$\sigma$ binned for plotting purposes
  only.
\label{nwxray}
}
\end{figure}

\begin{figure}
\includegraphics[width=15cm]{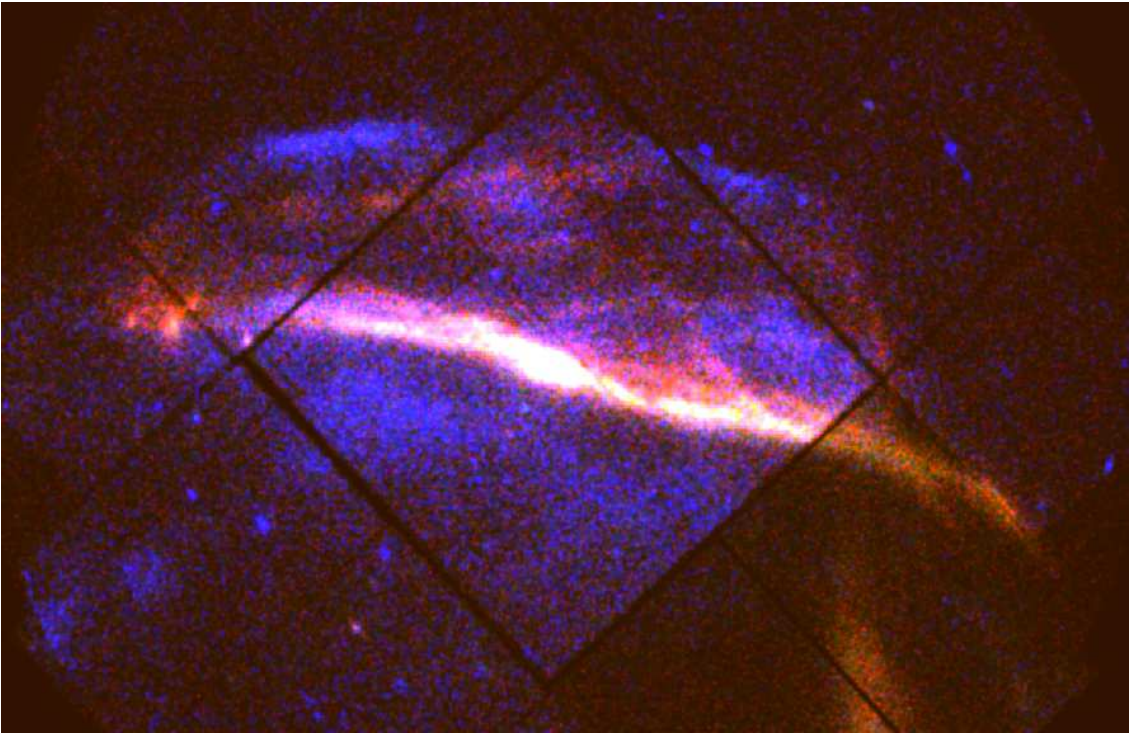}
\caption{EPIC-MOS image of the NW filament, with 0.3-0.7 keV emission
  in red, 0.7-1.3 keV emission in green, and 1.3-6.8 keV nonthermal
  emission shown in blue. The low-energy bands highlight thermal
  emission, while the high-energy band shows the location of
  non-thermal synchrotron radiation. Images have been smoothed with a
  2-pixel Gaussian to highlight extended emission.
\label{nw_nontherm}
}
\end{figure}

We show the 0.4-7 keV EPIC-MOS spectra from the NW in
Figure~\ref{nwxray}. The shock speed of this filament is 580-660 km
s$^{-1}$ \citep{ghavamian99}. Visible in the spectra is a continuum
extending to high energies, featureless except for an Fe K$\alpha$
line at $\sim 6.4$ keV. We account for these components with an {\it
  srcut} model for the synchrotron emission \citep{reynolds99} and an
Fe-only {\it vpshock} model with kT = 5 keV and $\tau_i$ = $10^9$
cm$^{-3}$ s. R02 also found Fe K$\alpha$ emission in the SW, where
shock speeds and plasma temperatures are also relatively low,
attributing it to reverse-shocked ejecta. It is interesting to note
that the nonthermal continuum in the NW does not show the same
morphology as the low-energy thermal emission. Thermal X-rays in that
region are well contained to the thin filament seen in
Figure~\ref{images}, but the synchrotron emission appears much more
diffuse. We show this in Figure~\ref{nw_nontherm}. Similar
discrepancies were observed in the SW, leading R02 to conclude that
the synchrotron emission is associated with reverse shocks driven into
the ejecta, and not the forward shock.

We follow the same modeling procedure as for the SW, and fit the
spectra with a {\it vpshock} model (with high-energy components added
as described above). In this region, we jointly fit MOS 1 and 2 data,
as this part of the remnant has not been observed with {\it
  Chandra}. The best fit model (reduced $\chi^{2}$ = 2.20) has kT =
0.28 (0.27, 0.30) keV and $\tau_i$ = 9.9 (8.7, 11.1) $\times 10^{10}$
cm$^{-3}$ s, with an emission measure (EM) for the thermal component
of 2.51 $d_{2.5}^2$ (1.90, 2.91) $\times 10^{57}$ cm$^{-3}$ (2.15
$d_{2.5}^2$ (1.63, 2.49) cm$^{-3}$ $\msun$). We let the abundances of
C, N (tied to C), O, Mg, Si, and Fe float freely in the model, and
find that all are in the neighborhood of solar (using \citet{wilms00}
abundances), with the exception of Si (1.7 solar) and Mg (0.6
solar). Ne was kept fixed to solar as an anchor for the other
abundances, as Ne is not expected in the ejecta of a Type Ia SN, and
is not depleted onto grains in the ISM. The overabundance of Si is
likely a result of using a single-temperature shock model to fit the
data. As we show in Section~\ref{hydro}, the shock structure is likely
more complicated than this, with varying shock speeds, and a
single-temperature shock model with kT = 0.28 keV could underpredict
the Si K$\alpha$ line at 1.8 keV. The model compensates for this by
making Si overabundant. Mg underabundance might be caused by
depeletion onto dust grains.

We find a post-shock density from matching IR data of $n_{H}$ = 2.0
(0.75, 5.0) cm$^{-3}$ and a radiating dust mass of 1.2 $d_{2.5}^2$
(0.33, 3.6) $\times 10^{-3}$ $\msun$. The large uncertainties for this
region are a result of the errors in the flux measurement (see
Section~\ref{fluxes}). The plasma temperature derived from this model
is significantly lower than that of \citet{vink06}, who fit {\it XMM}
spectra from a similar region in the NW with a two-component NEI
model, finding temperatures of $\sim 1$ and 4 keV. However, our
temperature is close to that reported by \citet{helder11}, although
they report lower ionization timescales. The difference is likely due
to the fact that our ``NW'' spectrum is extracted from a region that
is farther east than theirs, and much brighter in both X-rays and IR,
implying a higher density. Differences also result from different
spectral models used.

It is worth noting that the ionization timescales obtained from
fitting X-ray spectra, combined with the post-shock densities from
fitting IR data, allow a determination of the age of the shock. We
obtain ages of $\sim 700$ and $\sim 1300$ yr for the SW and NW,
respectively. These ages, derived from plane-shock model fits, are
underestimates for the true age of the remnant \citep{borkowski01b},
but do strongly favor a young SNR, strengthening the case for RCW 86
as the remnant of SN 185 A.D. Where dust is present and X-ray spectra
of the swept-up ambient medium can be examined, this method provides a
useful check on the ages of young SNRs.

In modeling dust emission from the post-shock environment, we focus
only on the {\it Spitzer} data, as the wavelengths covered with {\it
  Spitzer} are more appropriate for the bulk of the dust emission seen
in an SNR. WISE data at 22 $\mu$m do not further constrain the models
discussed later, and 12 $\mu$m data are sensitive to a
poorly-understood hot dust component, likely from very small dust
grains which are stochastically heated. The IR model above predicts a
flux of 100 mJy at 12 $\mu$m, while the measured flux from {\it WISE}
in this region is 430 mJy (albeit with a $\sim 40\%$ error). A similar
excess of short-wavelength IR emission above a model fit to 24 and 70
$\mu$m was found in our {\it Spitzer} analysis of Kepler's SNR
\citep{blair07}. Emission at this wavelength would likely come from
very small dust grains, and the physics of collisional heating and
sputtering of such small grains is not well understood. See Section
3.3 of \citet{blair07} for further discussion of the difficulties in
modeling short wavelength IR emission. However, this emission from
very small grains should have little effect on the overall properties
derived from the IR fits. If the 70 $\mu$m data were ignored, and only
the 12 and 24 $\mu$m data were fit with a single-temperature grain
model, the temperature of the grains would be 190 K, and the radiating
mass would be only 2.4 $d_{2.5}^2$ $\times
10^{-7}\ \msun$. Nevertheless, it is worth noting that significant
work still needs to be done to understand short-wavelength IR emission
from dust in SNRs.

\subsection{Dust-to-Gas Mass Ratio}

\citet{dwek87b} provide theoretical predictions for the ratio of the
total IR luminosity to the 0.2--4 keV X-ray luminosity, a quantity
that they refer to as the IRX ratio. At the temperature of the NW
region, this ratio should be $\sim 35$. Our dust model calculates the
amount of sputtering of grains in the post-shock gas, and for the NW
region we find that 19\% of the total dust mass has been
destroyed. Accounting for this mass, and considering only the
unabsorbed thermal component ({\it vpshock}) of our X-ray spectral fit
described above, we find an IRX ratio of 1.65, lower than the
predicted value by a factor of 20, which implies a dust-to-gas mass
ratio for the pre-shock ISM lower than the standard Galactic value
assumed by \citet{dwek87b}. The IRX ratio for the SW region is 1.4.

We can approach this issue in another way. The X-ray EM for the
thermal component in the NW region is 2.15 $d_{2.5}^2$ cm$^{-3}$
$\msun$. Since we have (from the IR fits) an independent determination
of the post-shock density, we can divide this out of the EM and get
the amount of gas that has been swept up by the shock, which is
roughly a solar mass. Normalizations to the IR spectra give the
radiating dust mass, and dividing these two gives another measure
(also distance-independent) of the dust-to-gas mass ratio, 1.37
$\times 10^{-3}$, where we have accounted for 19\% of the mass in
grains being destroyed via sputtering. The typical value for the
dust-to-gas mass ratio in the Milky Way is 7.5 $\times 10^{-3}$
\citep{weingartner01}. We thus confirm our finding from above of a low
IRX ratio, with a dust-to-gas mass ratio that is roughly a factor of 5
too low. The discrepancy between an IRX ratio that is 20 times too low
and a dust-to-gas mass ratio that is 5 times too low primarily results
from the assumption by \citet{dwek87b} that the plasma is in
collisional ionization equilibrium (CIE). In models for the X-ray
emission from RCW 86, the plasma is in a non-equilibrium ionization
state, which significantly raises the amount of X-ray emission when
compared to a CIE model. For example, if kT = 0.35 keV and $\tau_{i}$
= $7 \times 10^{10}$ cm$^{-3}$ s, a plane-shock model will produce
$\sim 3$ times more X-ray emission than a Raymond-Smith plasma in CIE.

We arrive at a similar conclusion in the SW, where the dust/gas
mass ratio is found to be 1.3 $\times 10^{-3}$, but note that both of
these methods will miss any dust that is too cold to emit at 70
$\mu$m. Similar dust/gas mass ratio shortfalls have been observed in
SNRs in both the Galaxy \citep{blair07,arendt10,lee09} and the LMC
\citep{borkowski06,williams06,williams11}.

\section{Progenitor Type of RCW 86}
\label{progenitor}

The nature of the progenitor SN of RCW 86 is uncertain, with most
authors favoring a CC origin. Support for the CC model primarily comes
from \citet{westerlund69}, who placed the location of the SNR within
an association of ten B-type stars at a distance of 2.5 kpc (although
one of these stars is likely a foreground object, and another has a
doubtful association based on its distance). These stars are spread
out over a much larger region of space than the periphery of the
SNR. Recent optical studies of the remnant have used line kinematics
of the narrow H$\alpha$ components of various shocks in the remnant,
along with Galactic rotation curves, to place the distance to the
remnant at 2.3-2.8 kpc \citep{sollerman03,rosado96}, confirming that
the remnant and the OB association are at approximately the same
distance.

However, this proximity to the association does not preclude a Type Ia
progenitor much older and less massive than the B-type stars found in
this region of our Galaxy.  This is illustrated by N103B, a young
remnant in the Large Magellanic Cloud known to be the result of a Type
Ia SN, that is apparently associated with the H II region
N103. \citet{badenes09} examined the local star formation history near
N103B and found a prominent extended peak between 100 and 50 Myr. A
mixture of young and old stellar populations is likely present near
RCW 86 as well, so both Type Ia and CC explosions may occur at this
particular location within our Galaxy.

Secondary support for the CC origin comes from the suggestion, first
made by \citet{vink97}, that RCW 86 is an explosion into a
cavity. Cavity explosions are typically associated with CC SNe and the
strong winds that some massive stars may blow off prior to
explosion. However, as shown in \citet{hachisu96} and
\citet{badenes07}, accreting WD systems, the progenitors of
single-degenerate Type Ia SNe, can also create substantial wind-blown
cavities.

Both the arguments in favor of a CC origin are valid, but are
circumstantial and based on general associations rather than specific
observations of the remnant. Here, we critically examine available
observations of RCW 86, and find that the remnant is more likely the
remnant of a Type Ia SN.

1. {\it Balmer-dominated shocks}. The entire periphery of the remnant
shows H$\alpha$ emission from non-radiative shocks encountering at
least partially neutral material \citep{smith97}. The mechanism for
this H$\alpha$ emission is described in \citet{chevalier80}. At the
densities derived above from IR observations, H recombination times
are $\sim 10^{5}$ yr. Most CC SNe ionize the surrounding medium in one
of two ways, either from the radiation from the shock breakout (in the
case of a red supergiant) or before the SN from the ionizing flux from
the progenitor star itself (in the case of a Wolf-Rayet star). A hot,
compact progenitor of a stripped CC Type Ib/c SN would produce
relatively few ionizing photons in the explosion, but would have a
strong ionizing flux before the explosion, as the progenitors of these
SNe are massive O-type stars during their main-sequence lifetime,
which proceed through a luminous blue variable (LBV) phase to shed
their H (and possibly He) envelope, becoming Wolf-Rayet stars. Less
massive ($< 25\ \msun$), cooler stars will be red supergiants (RSG) at
the time of the explosion. While they are too cool to ionize material
beforehand, their large size results in a large ionizing flux at shock
breakout. \citet{chevalier05} discusses the ionizing radiation from
the shock breakout of a CC SN. For the amount of mass that the shock
has encountered in RCW 86 (as constrained from model fits to IR data;
see Section~\ref{hydro}), explosions of RSGs (R $\sim 600 \rsun$),
including Type IIP, Type IIL, and even Type IIb SNe like 1993J
\citep{woosley94} are likely ruled out, as they produce more than
enough ionizing radiation. The progenitors of Type IIn SNe are
generally thought to be extremely massive ($> 80\ \msun$) O-type stars
which evolve through an LBV phase \citep{galyam07}, although
\citet{smith09} point out that LBVs may not be the only possible
progenitors of Type IIn SNe, and that they may arise from red
supergiants as well. In either case, the mass contained in the CSM is
enormous, and there is no evidence in RCW 86 for interaction of the
blast wave with a dense CSM. The only remaining CC scenario which
cannot be ruled out from this perspective is an explosion of a blue
supergiant (BSG) progenitor with a massive H envelope, such as SN
1987A, producing only a modest number of ionizing photons, but see
points 2 and 3 below.

2. {\it Fe abundance in ejecta}. Even if some CC models cannot be
ruled out from argument 1, all CC SNe produce large amounts of O with
respect to Fe in the ejecta \citep{woosley95}. There is no evidence
for super-solar O abundances anywhere in the remnant; both the spectra
from the NW and the SW are consistent with typical ISM abundances for
O. Fe L-shell emission is consistent with solar abundances as well
(see Table~\ref{xrayparams}), but Fe K$\alpha$ emission cannot be
explained by this model. The properties of the fits listed in
Table~\ref{xrayparams} indicate that we are seeing shocked ISM. The
6.4 keV line observed in both regions is most easily explained by a
high-temperature (5 keV), low $\tau_{i}$ ($\sim 1 \times 10^{9}$
cm$^{-3}$ s) plasma. As R02 and \citet{ueno07} point out, a plasma at
this temperature requires collisionless heating of electrons at the
reverse shock. An O-rich and Fe-rich plasma in such a low ionization
state does not produce L-shell emission from Fe, but H and He-like O
K-lines are expected to be prominent, even if the ionization age of
the plasma is lowered to a few $\times$ 10$^{8}$ cm$^{-3}$ s. A study
of X-ray emission in the SW using {\it Suzaku} found a lower limit of
0.05 $d_{2.5}^2$ $\msun$ of Fe from a region covering 7\% of the
remnant's surface area \citep{ueno07}. A similar study of the eastern
limb by \citet{yamaguchi08} found $> 0.06\ d_{2.5}^2$ $\msun$ of Fe
(with a possible upper limit an order of magnitude higher) from a
region covering 2.5\% of the remnant. As we show in
Section~\ref{hydro}, we find $3 \times 10^{-3} \msun$ in the NW, from
a region containing 1.5\% of the remnant.

Can oxygen simply be hidden? This would require a very low ionization
state of the gas. \citet{ueno07} derive their Fe mass listed above by
assuming that the electron density in the shocked ejecta is 0.1
cm$^{-3}$, and that the ejecta were shocked 1000 years ago. They
consider this electron density to be an upper limit, as anything
higher would raise the ionization timescale of the gas to levels ($> 3
\times 10^{9}$ cm$^{-3}$ s) that would result in strong Fe L-shell
lines. Lowering the density to the point where oxygen lines would be
weak, say, at $\tau_{i}$ = 10$^{8}$ would raise the amount of Fe in
the SW region by a factor of 30, to 1.5 $\msun$. While this amount of
Fe appears excessive even for a Type Ia SN, it would also rule out all
but the most exotic CC explosions. In the NE, \citet{yamaguchi08} fit
the centroid of the Fe K$\alpha$ line with Suzaku, and found
$\tau_{i}$ = 2.3 $\times 10^{9}$ cm$^{-3}$ s, a sufficient timescale
to produce strong O lines, yet none are reported.

3. {\it BSG Incompatibility with Observations}. \citet{gvaramadze03}
proposed that the remnant is the result of the explosion of a BSG,
moving through a bubble that the progenitor created during its main
sequence lifetime. A large amount of mass may have been expelled at
the RSG stage, facilitating transformation into the BSG. While the
large amount of Fe seems to rule out any CC progenitors, there are two
other arguments against this particular scenario. First, a slow wind
blown at the RSG stage is expected to be nitrogen-rich, as observed in
the CSM of SN 1987A \citep{mccray93}, but no nitrogen enhancement has
been reported \citep{leibowitz83}. Secondly, the explosion of a
massive blue supergiant, such as the progenitor of SN 1987A, leads to
the formation of an iron-nickel bubble \citep{li93}. As shown in
\citet{williams08}, an iron-nickel bubble forms dense clumps of O-rich
material, which should show strong optical or X-ray emission. Again,
no O-rich ejecta have ever been reported in this remnant. The failure
of this elaborate CC scenario demonstrates the difficulty in finding a
plausible CC model for RCW 86.

4. {\it Lack of central point source or pulsar-wind nebula}. While CC
explosions, such as Cas A \citep{pavlov00}, often leave behind a
compact remnant, Type Ia explosions should not.  \citet{fesen79}
conducted an optical survey of several known historical SNRs in the
Galaxy, finding no central point source in RCW 86 consistent with a
compact remnant down to a magnitude of 18.5. \citet{gvaramadze03}
report the possibility of a point source from a compact object in the
{\it Chandra} observations of the SW shell. However, this source is
located near the edge of the remnant, and if associated with the
supernova of 185 A.D., the transverse velocity of the source would be
in excess of 5000 km s$^{-1}$, even assuming an off-center
explosion. \citet{kaplan04} conducted a search with {\it Chandra} of
the interior of the remnant, finding no neutron star (down to a limit
of one-tenth the luminosity of the Cas A neutron star) or pulsar-wind
nebula. They assume the maximum reported distance of 2.8 kpc, and note
that their limits would be even more constraining if the remnant were
closer. However, their {\it Chandra} field of view does not cover the
entire remnant, and they are unable to rule out less likely compact
objects, such as massive neutron stars, neutron stars with exotic
interiors, and quiescent black holes.  \citet{kaspi96} find no radio
pulsar down to a limit of 1.3 mJy at 436 MHz and 0.2 mJy at 1520
MHz. No pulsar-wind nebula has ever been reported in the radio.

Despite the previously mentioned proximity to an OB association and
the suggestion that RCW 86 is the result of a cavity explosion, the
specific characteristics of the remnant are more easily explained by a
Type Ia explosion. We cannot definitively rule out all CC progenitors,
but we can say that any such event would be a very unusual CC SN. It
would have to produce significant amounts of Fe relative to O in the
ejecta, while being a weak source of ionizing radiation, either before
the SN or during the shock breakout phase. Its pre-supernova winds
would be inconsistent with the nitrogen-rich wind expected in the
transition from an RSG to a BSG, and any compact remnant left behind
would have to be less than one-tenth as luminous as the neutron star
in Cas A.

It is possible to confirm or refute the conclusion of RCW 86 being a
Type Ia SN. Light echoes from historical SNe can reflect the spectrum
of the SN event itself, and matching of detected echoes with SN
spectral templates has been used in several cases to determine the SN
type \citep{krause08}. Detection of a light echo spectrum from this
remnant that could be matched to a Type Ia template would confirm our
conclusion. A caveat to this method is that not all thermonuclear
explosions of white dwarfs look like Type Ia SN. A notable case of
this is SN 2002ic, where a strong H emission line was detected in the
spectrum of a thermonuclear SN \citep{wang04}. Alternatively,
detection of a compact remnant (neutron star, pulsar-wind nebula, or
black hole) that is definitively associated with the remnant would
clearly imply a CC explosion. Finally, although the remnant does not
belong to the ``O-rich'' class of SNRs (a group of approximately half
a dozen SNRs categorized by strong optical oxygen emission), it is
still possible that deep X-ray observations of the center of the
remnant could show oxygen in the ejecta.

\section{Hydrodynamic Modeling}
\label{hydro}

Perhaps the most puzzling aspect of RCW 86 is its large size, given
its likely association with the supernova of 185 A.D. At a distance of
2.5 kpc, the diameter of the remnant is $\sim 25$ pc, requiring an
average shock speed of $> 7000$ km s$^{-1}$. Measured shock speeds
vary greatly within the remnant. In the regions of interest considered
in this paper (where the shell is detected at both 24 and 70 $\mu$m),
shock speeds from optical spectroscopy are 500-700 km s$^{-1}$
\citep{rosado96,ghavamian01} , while shock speeds in the eastern limb
from the proper motion of X-ray emitting filaments have been reported
as high as 6000 ($\pm 2800)$ km s$^{-1}$ \citep{helder09}. The
detection of thermal X-ray emission at ionization timescales of 6 --
10 $\times 10^{10}$ cm$^{-3}$ s distinguishes this remnant from SN
1006, another young remnant of a Type Ia explosion that is very large
for its age. There, ionization timescales are shorter by nearly an
order of magnitude \citep{yamaguchi08b}.

\citet{badenes07} considered models of Type Ia explosions into bubbles
blown by accretion wind outflows from the progenitor system, noting
that most known Type Ia Galactic SNRs are inconsistent with this
scenario. However, they note that RCW 86 may be the exception to this,
and could in fact be the result of just such an event. The mechanism
for the creation of the winds modeled by \citet{badenes07} is
described in \citet{hachisu96}. A high mass accretion rate from a
red-giant (RG) or main-sequence companion onto a white dwarf (WD) can
lead to the formation of a large, loosely bound envelope surrounding
the WD. This envelope is optically thick, and continuum-driven winds
from the WD can eject several tenths of a solar mass from the
system. The wind stops when the mass transfer rate falls below $\sim 6
\times 10^{-7} \msun$ yr$^{-1}$, which can be as long as 10$^{5} -
10^{6}$ yr before the WD accretes enough mass to reach the
Chandrasekhar limit.

We perform hydrodynamic simulations of Type Ia explosions into various
surroundings using VH-1, a numerical hydrodynamics code developed by
J. Blondin and collaborators based on the piecewise parabolic method
of \citet{colella84}. We use the exponential ejecta density profile of
\citet{dwarkadas98}, and assume (except where otherwise noted) an
explosion energy of 10$^{51}$ ergs and an ejected mass of 1.4
$\msun$. Densities reported below are in units of either H atoms
cm$^{-3}$, where we use a scaling of $n_{0} =\rho_{0}/2.34 \times
10^{-24}$ gm cm$^{-3}$, appropriate for an ISM with a 10:1 H:He ratio,
or in amu cm$^{-3}$. We seek solutions which approximately reproduce
the known age, radius, and velocity. Radius is a difficult quantity to
determine for the somewhat unusual morphology of RCW 86. We chose a
point at the approximate geometric center of the remnant and measure
(assuming $D=2.5$ kpc) radii of 12 and 13 pc to the NW and SW
filaments, respectively. Since the NW represents the simpler geometry
of the two filaments, we chose 12 pc as the radius to be reproduced in
the hydro simulations.

\subsection{Explosion into a Uniform Medium}

We begin with the simplest case, an explosion into a medium with
constant density, as modeled in \citet{dwarkadas98}. In a 1-D hydro
model, assuming the standard explosion parameters above, the forward
shock can reach the desired radius in 1825 years by encountering a
constant density of $n_{0}$ = 0.033 cm$^{-3}$. In addition to the
density in this model being more than an order of magnitude below that
required to produce IR emission in the NW and SW, the current forward
shock velocity in this model is $v_{s}$ = 3000 km s$^{-1}$, well above
that measured from optical spectroscopy.

An alternative model can reach 12 pc with a current shock velocity of
600 km s$^{-1}$ by encountering a medium of $n_{0}$ = 0.75
cm$^{-3}$. While this model is consistent with both the measured shock
velocities and the inferred densities from IR, the time required for
this shock to reach the observed radius is 7000 years, ruling out an
association with SN 185 A.D. A third option is to require the shock to
decelerate to 600 km s$^{-1}$ in 1800 years, but this model requires
an ambient density of $n_{0}$ = 40 cm$^{-3}$ and places the current
radius of the forward shock at 3.2 pc. The distance to the remnant
derived from this value would be 650 pc, well inside any previous
distance estimates. Similar calculations show that reasonable under-
or over-energetic SNe likewise fail to reproduce anything close to the
observed remnant.

\subsection{Wind-Blown Bubble Model}

Given the results of models discussed above, we can rule out the case
of a blast wave from an explosion in 185 A.D. encountering a uniform
ambient density. The next level of complexity that we consider is a
low-density, uniform bubble (region I) surrounded by higher-density,
uniform ISM (region II). A constant wind blown from the progenitor
system would lead to an $r^{-2}$ density profile, but a uniform bubble
could be created by simply switching off the wind a few thousand years
prior to the explosion; delay times prior to explosion in
\citet{badenes07} are well in excess of this in most cases. Since our
IR modeling gives us a direct measurement of the post-shock density of
the material recently shocked, we have one more constraint to use in
these models, and we fix the ambient density in the ISM to the
post-shock density from IR fits ($\sim 2$ cm$^{-3}$) divided by a
compression ratio of 4, i.e. $n_{0,II}$ = 0.5 cm$^{-3}$. With this
number fixed, the number of free parameters in our bubble model is
only two: the density within the bubble, which we define as $n_{0,I}$,
and the location of the bubble ``wall,'' which we define as $\bar{R}$.

\begin{figure}
\includegraphics[width=15cm]{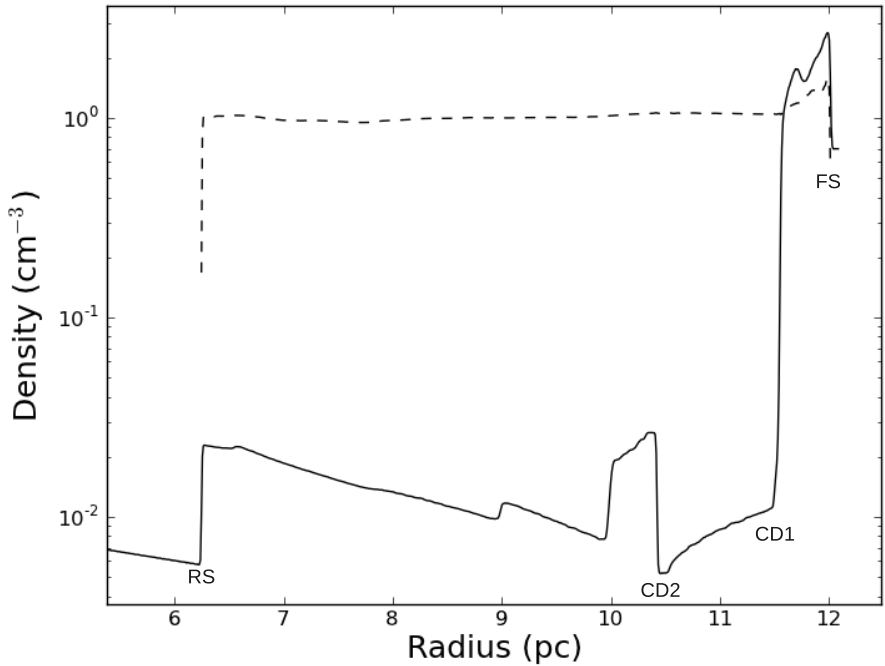}
\caption{Density profile of the wind-blown bubble model, as described
  in the text, in one dimension. Identified in the image are the
  forward shock (FS), the contact discontinuity separating the shocked
  ISM material from the shocked bubble material (CD1), the contact
  discontinuity separating the shocked bubble material from the
  shocked ejecta (CD2), and the reverse shock (RS). The collision of
  the blast wave with a shell of density contrast 250 leads to density
  structures from transmitted and reflected shocks, which can also be
  seen in the figure. Pressure, normalized to 1 at CD2, is shown as a
  dashed line.
\label{1dmodel}
}
\end{figure}

We find a good fit to the observed radius and age from a model with
$n_{0,I}$ = 0.002 cm$^{-3}$ and $\bar{R}$ = 10.8 pc. The current shock
speed in this model is 740 km s$^{-1}$, which roughly agrees with the
600-700 km s$^{-1}$ shock speeds measured from optical data in the NW
and SW, respectively. We show the density profile from a
one-dimensional (1-D) simulation of this scenario in
Figure~\ref{1dmodel}. In this model, the shock races through the
low-density bubble in 725 years, reaching the shell with a velocity of
8900 km s$^{-1}$. Upon hitting the shell of density contrast 250, the
transmitted shock speed into the dense material is 1300 km s$^{-1}$.
\citet{borkowski97} provide an analytic estimate for the transmitted
shock speed when transitioning from a bubble to a dense shell. If
$\delta$ is the density ratio, $n_{0,II}/n_{0,I}$, between the dense
shell and the bubble, then the pressure jump, $\beta \equiv
p_{2}/p_{1}$, is

\begin{equation}
\beta \cong 6(1+1.701\delta^{-1/2}\ - 0.4018\delta^{-1}\ +
0.2274\delta^{-3/2}\ - 0.0874\delta^{-2})^{-2},
\end{equation}

\noindent
and the transmitted shock speed is

\begin{equation}
V_{t} = (\beta/\delta)^{1/2} V_{B}
\end{equation}

\noindent
where $V_{B}$ is the speed of the shock when it reaches the wall. With
$\delta$ = 250 and $V_{B}$ = 8900 km s$^{-1}$, the transmitted shock
speed in this approximation is 1250 km s$^{-1}$.

The shock has been in the dense ISM for 1175 years, sweeping up a
total mass of $\sim 35$ $\msun$. The reverse shock, meanwhile, has
shocked virtually all of the ejecta (1.2 $\msun$). The ionization
timescale of the forward shocked material implied by this model,
$\tau_{i}$ = 8.9 $\times 10^{10}$ cm$^{-3}$ s, agrees quite well with
the values of 9.8 $\times 10^{10}$ cm$^{-3}$ s and 6.0 $\times
10^{10}$ cm$^{-3}$ s derived from X-ray fits to the NW and SW. We can
use a semi-analytic approximation to check the final shock speed (740
km s$^{-1}$), by assuming that the internal energy in the shocked
ejecta is the dominant energy component, significantly greater than
the internal energy in the shocked shell or the kinetic energies of
the ejecta and shell material. In this approximation, the forward
shock is completely driven by the pressure of the shocked ejecta, and
the shock speed is given by $V_{S} = (4p/3\rho_{S})^{1/2}$, where $p$
is the thermal pressure, $p = E/2\pi \bar{R}^{3}$. For an explosion
energy of 10$^{51}$ ergs and $\bar{R}$ = 10.8 pc, p = 4.3 $\times
10^{-9}$ ergs cm$^{-3}$. In this approximation, $V_{S}$ = 700 km
s$^{-1}$. In reality, the internal energy of the shocked ejecta,
obtained by numerically integrating the output of the hydro model, is
$\sim 40\%$ greater than the sum of the kinetic and internal energies
of the shocked ISM, so this approximation is not entirely valid, but
the general agreement is encouraging.

Once the assumption of a uniform ambient density is abandoned, the
parameter space for possible density structures that the shock can
encounter is vast. The solution above of a shock running through a
low-density bubble and encountering the much higher density ISM is not
unique, but it is not our aim to explore every possible scenario. This
model represents the simplest solution consistent with the available
observed constraints. It is interesting to note that while we do {\it
  not} use the models of \citet{badenes07} as a constraint in any way,
the best-fit model listed above looks remarkably like model H1 from
their paper, which consists of a white dwarf of 1 $\msun$ in a binary
system with a donor star of 2 $\msun$, with an initial period for the
system of 2 days. The total mass lost by the system is 0.15 $\msun$,
with an outflow velocity of 1000 km s$^{-1}$, and the time between the
onset of the final mass transfer and the SN explosion is 5 $\times
10^{5}$ yr. The ``bubble'' created by this system has a roughly
constant density of 0.001 cm$^{-3}$ with a shell wall located at $\sim
17$ pc. The density in the ISM is fixed at 0.42 cm$^{-3}$, with the
exception of a shell at the bubble-ISM interface with a density higher
by a factor of $\sim 2$. We assume that the blast wave is contained
within this shell.

\begin{figure}
\includegraphics[width=14cm]{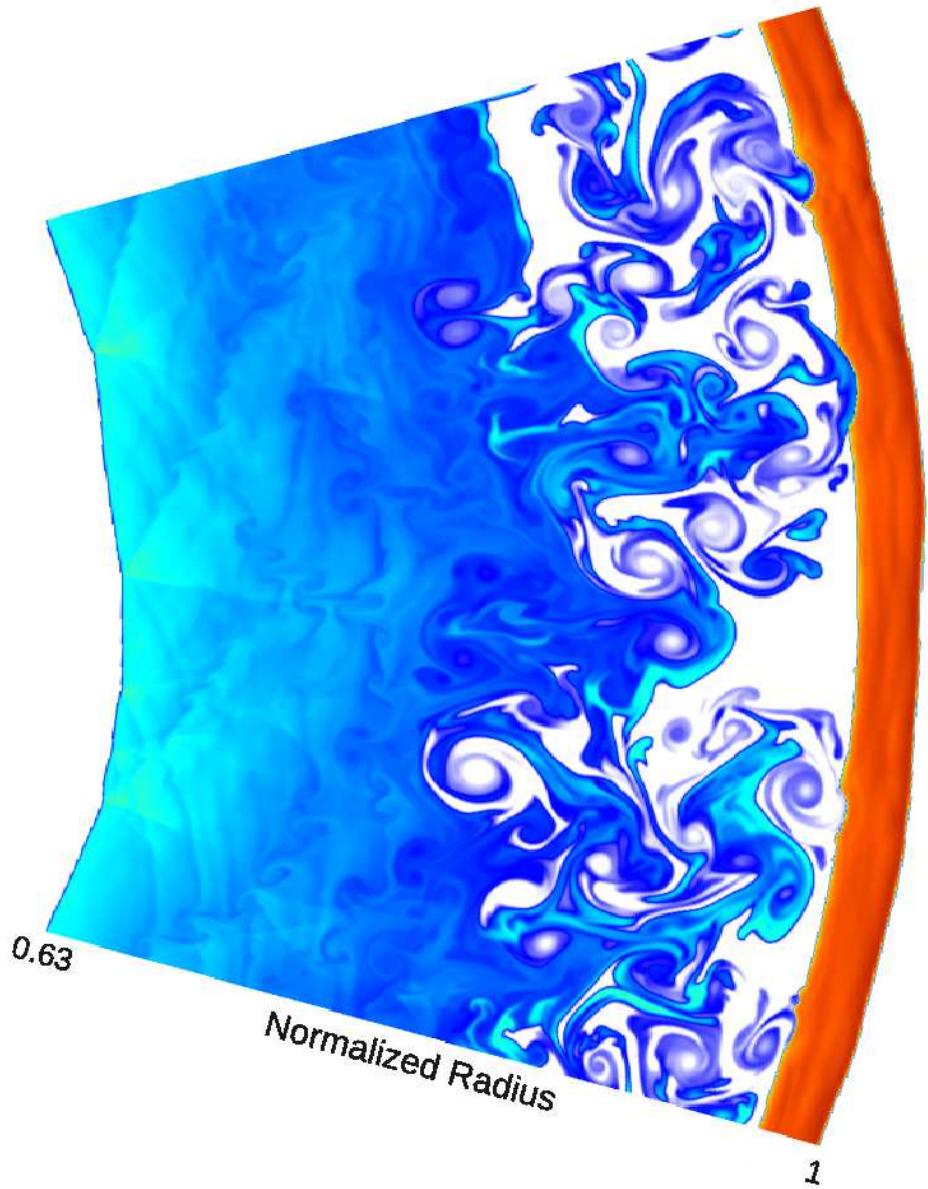}
\caption{Two-dimensional density structure in the wind-blown bubble
  model for an explosion located at the center of the bubble. Radius
  is normalized to 1 at the forward shock location ($\sim 12$
  pc). Marked radii are forward shock (1) and reverse shock
  (0.63). Orange and red colors indicate shocked shell material, white
  represents shocked ``bubble'' material, and blues represent
  reverse-shocked ejecta. The Rayleigh-Taylor unstable contact
  discontinuity is traced by the border between white and blue. The
  collision with the dense shell brings the contact discontinuity much
  closer to the forward shock than is possible from a shock expanding
  into a uniform medium. Polar angle from 0.4$\pi$ to 0.6$\pi$.
\label{2dmodel}
}
\end{figure}

This model also provides an explanation for the Fe K$\alpha$ line seen
in the X-ray spectrum from the narrow extraction region in NW, where
the elliptical region is quite narrow compared to the shock
radius. The extraction region, shown in Figure~\ref{images}, begins at
a radial distance of 93\% of the forward shock radius. Shocked ejecta
cannot exist this far out in the 1-D model, but this is only
indicative of the limitations of modeling a multi-dimensional system
in 1-D. In two dimensions (2-D), the instabilities generated by the
interaction of the blast wave with the density profile in the bubble
model allow the ejecta to pile up quite close to the forward shock. We
show the output of a 2-D model at the current age of the remnant in
Figure~\ref{2dmodel}, which shows significant amounts of shocked
ejecta at a radius of $> 93\%$ of the forward shock radius.

We can also use the output of the hydrodynamic modeling, combined with
X-ray spectroscopy, to estimate the amount of reverse-shocked Fe
present in the NW extraction region. The EM, as fit to the Fe
K$\alpha$ line in the {\it XMM-Newton} spectrum for the Fe-only
component, is 7.2 $\times 10^{-6}$ cm$^{-3}$ $\msun$. The hydro models
give the density in the reverse shocked ejecta, which over the small
range of interest ($>$ 93\% of the forward shock radius) is roughly
constant at 1.38 $\times 10^{-2}$ amu cm$^{-3}$. If this is all Fe,
then $n_{Fe}$ = 2.5 $\times 10^{-4}$ cm$^{-3}$. If we assume 10 free
electrons per Fe atom (to avoid ionizing Fe to the point of producing
Fe L-shell emission), then an electron density of 2.5 $\times 10^{-3}$
cm$^{-3}$ leads to an Fe mass of 3 $\times 10^{-3}$ $d_{2.5}^2$
$\msun$, a lower limit given the approximations used. This is a
relatively small amount of iron, but the size of the extraction region
for the NW covers only about 1.5\% of the surface area of the
remnant. As we mention in Section~\ref{progenitor}, similar
calculations by \citet{ueno07} and \citet{yamaguchi08} find a lower
limit of 0.11 $d_{2.5}^2$ $\msun$ of Fe from less than 10\% of the
area of the remnant.

\subsection{Off-Center Explosion}

Despite the general agreements listed above between our 2-D
hydrodynamic model and the measured properties of the remnant, one
thing this model cannot account for is the extremely high shock speed
reported for the NE limb, found from {\it Chandra} proper motions to
be 6000 $\pm 2800$ km s$^{-1}$ \citep{helder09}. Our hydro model
assumes spherical symmetry for the distance from the explosion to the
shell wall, and this symmetry must be broken to explain shock speeds
that differ by an order of magnitude from one side of the remnant to
the other. It is beyond the scope of this paper to provide a detailed
multi-dimensional model to explain the morphological ``squareness''
and varying shock speeds of RCW 86; such a model would undoubtedly be
highly complex. Rather, we seek the simplest case that can broadly
explain the observations, that of an off-center explosion into a
wind-blown bubble, in which the shock has not yet reached the bubble
wall in the NE.

\begin{figure}
\includegraphics[width=15cm]{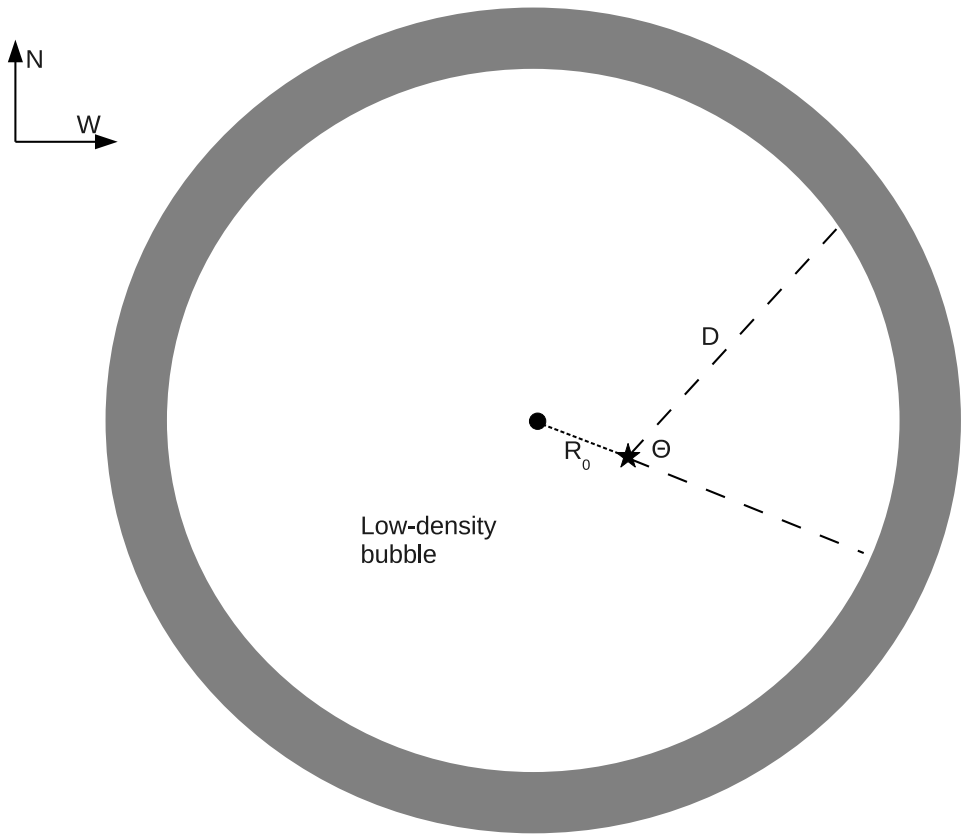}
\caption{Geometry of off-center explosion model, as described in the
  text. The center of the spherically symmetric bubble is marked by a
  dot, and point of explosion is marked by a star. In this model, the
  radius from the center of the bubble to the inner wall of the shell
  is the approximate observed radius of the remnant, 12 pc. $R_{0}$ is
  the distance from the center of the bubble to the center of the
  explosion, which is variable in this model.
\label{geom}
}
\end{figure}

We assume that the wind-blown bubble is symmetric with a radius of 12
pc. We then vary the distance, R$_{0}$, from the center of the bubble
to the center of explosion. R$_{0}$ is the only free parameter in this
model; we assume the same bubble and shell densities as above
(n$_{0,I}$ = 0.002 cm$^{-3}$ and n$_{0,II}$ = 0.5 cm$^{-3}$). In this
model, the distance, D, from the center of explosion to any point on
the wall of the shell is given by

\begin{equation}
D(R_{0},\theta) = (R_{0}^{2} - 2R_{0}R \cos\ \theta + R^{2})^{\frac{1}{2}},
\end{equation}

\noindent
where $R$ is the fixed radius of the shell wall (in this case, 12 pc)
and $\theta$ is the polar angle between the center of explosion and a
point on the shell wall. A diagram of the geometry of this model is
shown in Figure~\ref{geom}. In 2-D, we model this as a semi-circle
from 0 to $\pi$ radians, using a spherical geometry with reflecting
boundary conditions at 0 and $\pi$.

\begin{figure}
\includegraphics[width=17cm]{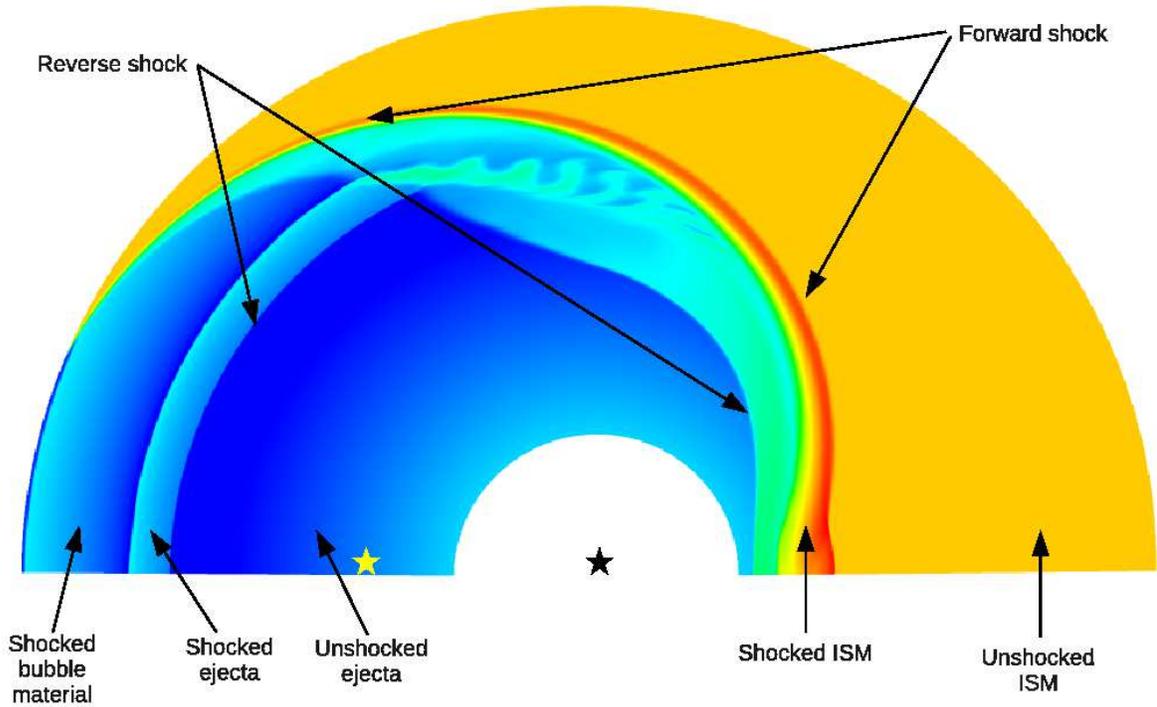}
\caption{Density structure in the off-center explosion model, as seen
  in the plane of the sky, shown at the current age of the
  remnant. The center of the explosion is marked with a black star,
  and the center of the bubble is marked with a yellow star. The
  distance between the two ($R_{0}$ in the text) is 7 pc. At $\theta =
  0$, measured from the center of explosion, the foward shock has
  reached a radius of 13.5 pc with respect to the center of the
  bubble, while at $\theta = \frac{\pi}{2}$, the shock is at 12.7 pc,
  again with respect to the center of the bubble. On the far side
  ($\theta = \pi$), the forward shock is still in the low-density
  bubble, and has not yet reached the wall. Lack of an instability at
  the contact discontinuity is the result of the low angular
  resolution used in this model.
\label{offcenter}
}
\end{figure}

Within this symmetric shell model, a significant offset of the
explosion from the center of the bubble is required for a shock that
roughly reproduces the observables (shock velocity, density, age) in
the SW, yet still allows for the extremely high shock speeds in the
NE. To avoid hitting the far side ($\theta = \pi$) of the shell wall
in 1825 yr, $R_{0}$ must be at least 7 pc. Assuming this value, the
shock encounters the shell wall on the near side ($\theta = 0$; the SW
corner of the remnant) in $\sim 300$ yr, and reaches a final radius
(with respect to the center of the remnant) of 13.5 pc with a velocity
of 1000 km s$^{-1}$. The hydrodynamic output of this model is shown in
Figure~\ref{offcenter}. \citet{rosado96} estimated the shock velocity
in the SW region to be 800 km s$^{-1}$. At $\theta = \pi/2$, which
would roughly correspond to the location of the NW filament, the shock
hits the filament in $\sim 1000$ yr, reaching a final radius (with
respect to the center of the remnant) of 12.7 pc with a velocity of
550 km s$^{-1}$, numbers which agree quite well with
observations. This model implies that the shock has only recently
encountered the shell in the E and SE. The ionization timescale of the
shocked ISM in the E, as reported by \citet{yamaguchi08}, was 7.7
$\times 10^{10}$ cm$^{-3}$ s. This is rather high for a recently
shocked shell, but the authors point out that densities in this region
are likely much higher than the rest of the shell, and that the
ionization timescale of the reverse shocked Fe implies a shock age of
$< 380$ yr. Thus, there is a broad consistency of observable
parameters with the off-center explosion model, including the
transition from thermal to nonthermal X-rays in the NE corner of the
remnant, where, in our model, the shock has not yet hit the bubble
wall. Getting the progenitor system away from the center of the bubble
does not require excessively high velocities. If the delay time
between the onset of the fast wind and the SN explosion is as little
as 10$^{5}$ years, a velocity of 65 km s$^{-1}$ would cause an offset
of 7 pc. Delay times of $10^{6}$ years or more require velocities of
only a few km s$^{-1}$.

\section{Synchrotron X-rays}

The nonthermal X-ray emission from RCW 86, especially in the SW,
differs in some important ways from that seen in other young SNRs.
Most young SNRs show synchrotron X-rays concentrated in ``thin rims''
at the periphery (see Reynolds 2008 for a review).  These rims are
commonly assumed to be limited in thickness by the distance an
electron can advect with the post-shock flow in a synchrotron loss
time (Vink \& Laming 2003; Parizot et al.~2006).  However, in RCW 86,
most synchrotron emission is more diffuse, except for the E and NE
edges.  The spectrum in the SW can be well described by emission from
a power-law electron distribution with exponential cutoff (model srcut
in XSPEC) (R02).  The characteristic ``rolloff'' frequency $\nu_{\rm
  roll}$ (critical, not peak, frequency of electrons with the
e-folding cutoff energy $E_m$) is about $10^{17}$ Hz (R02), implying
$E_m = 40(B_{10})^{-1/2}$ TeV, where $B_{10} \equiv B/10\ \mu{\rm
  G}$. R02 showed that, given a geometric effect that can speed up
particle acceleration in perpendicular shocks (Jokipii 1987), shocks
of the speed of the optical nonradiative shocks in the SW ($\sim 600$
km s$^{-1}$) could accelerate electrons to this energy. However,
without this obliquity effect, shock velocities in excess of 1000 km
s$^{-1}$ are more typically required.  Such shocks have not been
present in the SW since the collision about 1500 yr ago of the blast
wave with the bubble wall.  However, at that time the blast-wave speed
was of order 10,000 km s$^{-1}$ in our scenario, and acceleration to
very high energies (100 TeV or above) would have been possible due to
the strong dependence of maximum energy on shock speed ($E_{\rm max}
\propto V_s^2$ for loss-limited acceleration).  In this case,
presumably a ``thin rim'' morphology would have resulted, but the
rolloff frequency would have been considerably higher than observed
today.  But electrons with energies below $E_m$ of that time have
longer lifetimes: the time for an electron of energy $E$ TeV to lose
half its energy is $t_{1/2} = 1.3 \times 10^5 E^{-1} B_{10}^{-2}$
yr. Combining this with the relation between energy and characteristic
frequency, we find that electrons with characteristic frequency
$\nu_{\rm roll}$ can survive for a time $t$ if they radiate in a
magnetic field less than
$$B_{\rm max}(t) = 29 \left(t \over 1000 \ {\rm yr}\right)^{-2/3} 
\left(\nu_{\rm roll} \over {10^{17} \ {\rm Hz} }\right)^{-1/3} 
\ \mu{\rm G}.$$ 
(We note that if $B < 3.3 \ \mu$G, the magnetic-field strength with
energy density of the cosmic microwave background (CMB), then
inverse-Compton losses from CMB photons will dominate synchrotron
losses.  However, this is unlikely to be the case in RCW 86 or other
young SNRs.)  For RCW 86, electrons with $\nu_{\rm roll} \sim 10^{17}$
Hz can survive for 1500 yr if $B < 22 \ \mu$G.

Synchrotron emission from electrons accelerated over 1500 years ago
would not be expected to be concentrated in thin rims; it could occupy
the entire region of shocked bubble material, and would appear in
projection as diffuse emission extending back into the remnant
interior, as observed.  For younger remnants like Tycho, thin rims
(according to the synchrotron-loss scenario) imply much higher
magnetic fields which would deplete all electrons capable of producing
X-rays.  However, in RCW 86 and perhaps other cavity explosions, it is
natural to expect diffuse, long-lasting synchrotron X-ray emission, as
long as magnetic field strengths remain relatively low.  Acceleration
to very high energies, while the blast wave is still inside (or
within) the bubble, would produce a population of electrons that could
go on emitting a synchrotron spectrum up to X-ray energies, even if
the subsequent shock transmitted into the dense surroundings rapidly
becomes too slow to accelerate electrons to TeV energies.

A detailed analysis of the X-ray synchrotron emission from all around
the periphery of RCW 86 is beyond the scope of this paper, but should
be undertaken to test this suggestion and to clarify the current state
and past history of electron acceleration in RCW 86.

\section{Gamma-Ray Emission}
\label{gammaray}

We can use our IR fitted densities to make an order of magnitude
estimate regarding the gamma-ray emission observed. In the hadronic
model of gamma-ray emission, cosmic-rays escaping the shock collide
with thermal gas particles, producing $\pi^{0}$ particles which decay
into gamma-rays. The 1-10 TeV spectrum from H.E.S.S. is fit with a
photon index of $\Gamma$ = 2.41 \citep{aharonian09}. If extrapolated
back to 100 MeV, this gives a luminosity, L$_{\gamma, 100 MeV}$ = 1.19
$\times 10^{39}$ photons s$^{-1}$. Using the gamma-ray emissivity from
\citet{drury94}, the energy density in relativistic particles for a
spherical SNR required to produce a $\gamma$-ray luminosity of
L$_{\gamma}$ by this process is

\begin{equation}
u_{rel} = 2.12 \times 10^{-17} \left(\frac{L_{\gamma, 100
    MeV}}{10^{39}\ {\rm ph}\ {\rm s}^{-1}}\right)
\left(\frac{d}{2.5\ {\rm kpc}}\right)^{-3}
\left(\frac{\theta_{rad}}{1\ {\rm arcmin}}\right)^{-3}
\left(\frac{n_{H}}{{\rm cm}^{-3}}\right)^{-1} q_{\gamma}^{-1}\ {\rm
  ergs}\ {\rm cm}^{-3}
\end{equation}

\noindent
where L$_{\gamma, 100 MeV}$ is the luminosity, in photons s$^{-1}$, of
100 MeV $\gamma$-rays, d is the distance to the remnant,
$\theta_{rad}$ is the radius of the remnant in arcminutes, $n_{H}$ is
the post-shock proton density, and q$_{\gamma}$ is the production rate
of $\gamma$-rays, in units of s$^{-1}$ erg$^{-1}$ cm$^{3}$
H$^{-1}$. We take a representative value from \citet{drury94} of
q$_{\gamma}$ = 5 $\times 10^{-14}$ and calculate $u_{rel}$ = 5.1
$\times 10^{-8}$ ergs cm$^{-3}$ for RCW 86. If we make the (admittedly
crude) assumption that the NW filament typifies the entire remnant,
then the total energy density ($\rho_{0} v_{s}^{2}$) of the thermal
gas is 4.2 $\times 10^{-9}$ ergs cm$^{-3}$. In this case,
$u_{rel}/\rho_{0} v_{s}^{2}$ = 12, requiring an order of magnitude
more energy going into relativistic particles than is available in the
shock. A more realistic value for this ratio, 0.1, requires a
flattening of the gamma-ray spectrum below 1 TeV to $\Gamma$ = 1.9. We
thus confirm the conclusions of \citet{aharonian09}, who based their
argument for the low-energy hardening of the $\gamma$-ray spectrum on
the total energy from the SN of $10^{51}$ ergs.

Any particle acceleration occuring in the remnant would rob the
post-shock gas of energy, raising the shock compression ratio across
the shock. We cannot determine this ratio directly, since we do not
have a direct measure of the emitting volume of the NW
region. However, since we have the post-shock density and total
shocked gas mass, the volume for a ``standard'' shock with a
compression ratio of 4 would be 1.52 $\times 10^{57}$
cm$^{3}$. Assuming $D = 2.5$ kpc, the projected radii of the
elliptical region are 4.7 and 0.62 pc for the major and minor axes,
and the line-of-sight depth through the filament would be 4.3 pc, a
factor of 7 greater than the radial width. This is the minimum
line-of-sight, as anything shorter would imply a CR $<$ 4.

\section{Concluding Remarks}

{\it Spitzer} and {\it WISE} mid-IR images show the complete shell of
RCW 86, with the dominant source of emission being warm dust grains
heated and sputtered in the post-shock environment. A combined
IR/X-ray analysis yields post-shock densities in the non-radiative
regions of the NW and SW of $n_{H}$ = 2.0 cm$^{-3}$ and 2.4 cm$^{-3}$,
respectively, with $\sim 20$\% of the dust in these regions destroyed
via sputtering. Based on the IR/X-ray flux ratio, we find that the
dust/gas mass ratio in the pre-shock medium is lower by a factor of
$\sim 5$ from that typical for the Galaxy. The post-shock gas
densities derived from IR observations place strong constraints on the
total amount of mass that has been encountered by the shock, and favor
a young age for the SNR, strengthening the case for an association
with SN 185.

X-ray and optical evidence points to a Type Ia origin for the
progenitor of RCW 86. The presence of Balmer-dominated shocks around
the periphery of a young SNR means that the blast wave is encountering
neutral material, and most CC SNe ionize significant portions of the
surrounding ISM, much more than is observed here. There are SNRs which
are believed to be the result of explosions within a cavity and show
Balmer emission at the shock front (e.g., the Cygnus Loop, believed to
be the remnant of a CC SN), but these are older objects where the
forward shocks have encountered significantly more mass (the swept
mass of the Cygnus Loop is likely $> 100 \msun$, based on the
densities obtained by \citet{sankrit10} and \citet{raymond03}). Not
all CC explosions can be ruled out because of the H$\alpha$ emission,
but the amount of Fe in the ejecta relative to O and other elements
also indicates a Type Ia origin. The X-ray analyses of the amount of
Fe relative to O in the ejecta have been taken from references already
in the literature, but the constraints on the ionizing flux from the
progenitor come from the new IR observations presented here. There is
additional evidence to support the claim that RCW 86 is the result of
a Type Ia SN, in that blue supergiant explosions, which may avoid
ionizing the surrounding medium, are incompatible with optical
observations, and that no compact remnant or pulsar-wind nebula has
ever been detected.

Most intriguingly then, we find that if RCW 86 is the remnant of 185
A.D., and is also the result of a Type Ia SN, then the only
self-consistent model to explain the IR, X-ray, and optical
observations is that of an explosion into a cavity created by the
progenitor system, a model which requires a single-degenerate
progenitor. We used one and two-dimensional hydrodynamic models to
arrive at this conclusion, finding that a simple model of an explosion
in a low-density bubble surrounded by a higher density shell can
approximately reproduce observed shock radii, velocities, and
post-shock gas densities. The density contrast between the bubble and
the shell is 250. This model does not require an appeal to any special
circumstances, such as an exceptionally recent encounter with the
shell wall or a sub/super-energetic explosion. The high proper motion
of the X-ray emitting filaments in the NE, and thus the inferred high
shock speed there, requires the explosion to be off-center, so that
the shock in that region has not yet encountered the dense shell. In
this model, the blast wave began encountering the dense ISM $\sim
1500$ yr ago and continues to impact parts of the shell.

The fast, synchrotron X-ray emitting shocks in the NE are still
encountering ionized material within the bubble, and thus no Balmer
emission should be expected from these shocks. In the places where the
shock is encountering the shell of the bubble, it does so at an
oblique angle. X-ray proper motion measurements are sensitive to the
speed at which the shock moves as it encounters the shell, which is
$\sim 5500$ km s$^{-1}$ in the 2-D hydrodynamic model. The broad
H$\alpha$ line width, on the other hand, will reflect the post-shock
temperature within the dense shell, where the transmitted shock is
considerably slower. In our hydro model, the shock speed of the
H$\alpha$ emitting shocks within the dense shell is $\sim 800$ km
s$^{-1}$. We did not adjust this model to match the values for the NE,
nonetheless, these values are quite close to the 6000 and 1100 km
s$^{-1}$ shocks reported for that region by \citet{helder09}.

Thus, this model offers an alternative explanation for the apparent
discrepancy between the shock speed measured by proper motions and
that derived from optical line kinematics, which \citet{helder09}
attributed to significant cosmic-ray accleration, with $> 50\%$ of the
shock energy going into cosmic rays. While we do not rule it out, our
model does not require any acceleration of cosmic rays at the shock
front.

The synchrotron emission in RCW 86 is complicated, with the SW and NW
showing more diffuse emission than the NE, where rims are sharper. The
bubble model offers a natural explanation for this. In the regions of
the remnant where the shock encountered the shell long ago, the
electrons were accelerated by much faster shocks than are currently
present there. For reasonable values of the magnetic field, these
electrons can survive to the current age of RCW 86.

This idea may find applicability in another older SNR traditionally
ascribed to a cavity explosion: the Cygnus Loop, where GeV emission
has recently been detected by the {\sl Fermi} satellite (Katagiri et
al., submitted).  Electrons accelerated early in the life of the
Cygnus Loop, while its blast wave moved rapidly through the
low-density bubble, could persist for the estimated age of order
10,000 yr (e.g., Levenson et al.~1998) with energies as large as
$10\ B_{10}^{-2}$ TeV.  Such electrons could produce gamma rays by
inverse-Compton scattering of various photon fields.  As a rough
example, 1 GeV photons could be produced from upscattering cosmic
microwave background (CMB) photons ($E \sim 3 \times 10^{-4}$ eV) by
electrons with Lorentz factors $\gamma \sim (h\nu_{\rm out}/h\nu_{\rm
  in})^{-1/2} \sim 2 \times 10^6$, or electron energies of about 1
TeV.  Electrons with these energies could easily survive from an early
period of rapid particle acceleration as long as $B < 25\ \mu$G or so
in the regions they inhabit.  In a current magnetic field $B_{10} \sim
1$, they would have characteristic frequencies of synchrotron
radiation $\nu_c \sim 2 \times 10^{14}$ Hz, i.e., in the near-IR, so
no nonthermal X-rays would be expected.

Thus, RCW 86 may indeed be the first example of a Type Ia explosion in
a system in which the progenitor carved a wind-blown bubble, and can
provide new insight into the outflows from accreting systems. While
the hydro simulations themselves {\it cannot} distinguish between a
Type Ia and CC progenitor system, they can rule out the case of a
uniform ambient density. We find a general agreement with the
parameters of the bubble model necessary to explain the dynamics of
the remnant and the simulations of \citet{badenes07}, who modeled the
outflows of single-degenerate Type Ia progenitor systems. In such a
system, the companion star would retain its generally high velocity
after the explosion. A star in a binary system with a period of a few
days can have an orbital velocity, $V$, of up to several hundred km
s$^{-1}$. As shown in the recent work by \citet{kerzendorf09},
high-resolution spectroscopy to determine the rotational period can
test the viability of a potential companion star. The maximum angular
distance that a companion star could travel from the center of
explosion in 1825 yr is 47 $V_{300}\ d_{2.5}$ arcsec, where $V_{300}$
is the velocity of the companion in units of 300 km s$^{-1}$. Although
the center of explosion is currently unknown, it should be possible to
identify the companion star from a single-degenerate Type Ia
SN. Another promising possibility would be to find the light echo of
RCW 86 and obtain a spectrum of the SN itself.

\acknowledgments

We acknowledge support from {\it Spitzer} Guest Observer Grants JPL
RSA 1378047, NSF Theory Grant AST-0708224, and NASA Astrophysics Data
and Analysis Program Grant NNX11AB14G. This work is based [in part] on
observations made with the Spitzer Space Telescope, which is operated
by the Jet Propulsion Laboratory, California Institute of Technology
under a contract with NASA. This research has made use of software
provided by the Chandra X-ray Center (CXC) in the application package
CIAO. Support for this work was provided by NASA through an award
issued by JPL/Caltech.

\newpage
\clearpage

\begin{deluxetable}{lcc}
\tablecolumns{3}
\tablewidth{0pc}
\tabletypesize{\footnotesize}
\tablecaption{X-ray Spectral Models}
\tablehead{
\colhead{Model Parameters} & NW & SW}

\startdata

$N_{H}$ ($\times 10^{21}$ cm$^{-2}$) & 6.25$^{6.6}_{5.8}$ & 6.53$^{6.9}_{6.1}$\\
$kT$ (keV) & 0.28$^{0.30}_{0.27}$ & 0.39$^{0.42}_{0.37}$\\
$\tau$ ($\times 10^{10}$ cm$^{-3}$ s) & 9.9$^{11.1}_{8.7}$ & 6.0$^{6.6}_{5.5}$\\
C, N, O & 0.89$^{0.95}_{0.82}$ & 1.12$^{1.22}_{1.04}$\\
Mg & 0.57$^{0.61}_{0.52}$ & 0.63$^{0.66}_{0.60}$\\
Si & 1.72$^{2.04}_{1.40}$ & 1.20$^{1.32}_{1.07}$\\
Ca, Fe, Ni & 0.77$^{0.81}_{0.71}$ & 0.93$^{0.99}_{0.88}$\\
EM ($\times 10^{56}$ cm$^{-3}$) & 25.1$^{29.1}_{19.0}$ & 8.44$^{9.70}_{6.94}$\\
$\alpha$ & $\equiv$ 0.6 & $\ldots$\\
Log $\nu_{roll-off}$ (Hz) & 16.85$^{17.0}_{16.72}$ & $\ldots$\\
F$_{1\ GHz}$ (Jy) & 0.46$^{0.72}_{0.34}$ & $\ldots$\\
Reduced $\chi^{2}$ & 2.21 & 3.05\\
L$_{X}$ ($\times 10^{34}$ ergs s$^{-1}$) & 11.5 & 6.4\\

\enddata \tablecomments{Errors quoted correspond to 90\% confidence
  intervals. Abundances are from \citet{wilms00}. Models above do not
  include Fe-only component to account for Fe K-$\alpha$ line at 6.4
  keV. {\it srcut} parameters ($\alpha$, $\nu_{roll-off}$, \&
  F$_{1\ GHz}$) are frozen to a fit performed only from 2-6
  keV. Errors listed are for that fit alone. L$_{X}$ from 0.2-4 keV}
\label{xrayparams}
\end{deluxetable}  

\newpage
\clearpage

\begin{deluxetable}{lcc}
\tablecolumns{3}
\tablewidth{0pc}
\tabletypesize{\footnotesize}
\tablecaption{IR Models}
\tablehead{
\colhead{IR Measurements and Modeling} & NW & SW}

\startdata
$F_{24}$ (Jy) & 0.69$^{0.88}_{0.50}$ & 0.37$^{0.40}_{0.34}$\\
$F_{70}$ (Jy) & 3.10$^{4.16}_{2.04}$ & 1.41$^{1.49}_{1.33}$\\
$F_{70}/F_{24}$ & 4.5$^{6.5}_{2.5}$ & 3.8$^{4.1}_{3.5}$\\
$n_{H}$ & 2.0$^{5.0}_{0.75}$ & 2.4$^{2.75}_{2.1}$\\
$n_{e}$ & 2.4$^{6.0}_{0.90}$ & 2.9$^{3.3}_{2.5}$\\
M$_{dust}$ ($\times 10^{-4} \msun$) & 12.3$^{36.0}_{3.3}$ & 3.2$^{3.9}_{2.7}$\\
\% destroyed & 19 & 18\\
L$_{IR}$ ($\times 10^{34}$ ergs s$^{-1}$) & 19 & 9.1\\
M$_{gas}$ ($\msun$) & 0.89$^{2.4}_{0.36}$ & 0.24$^{0.29}_{0.19}$\\
M$_{dust}/$M$_{gas}$ & 1.38$^{1.5}_{0.92}$ $\times 10^{-3}$ & 1.3$^{1.7}_{0.94}$ $\times 10^{-3}$\\
IRX ratio & 1.65 & 1.4\\
Shock Age (yr) & 1310$^{3900}_{460}$ & 660$^{770}_{550}$\\

\enddata 
\tablecomments{M$_{dust}$ is pre-shock dust mass encountered
  by blast wave, i.e., has already been corrected for destruction of
  grains via sputtering. L$_{IR}$ is total IR luminosity, from 1-400
  $\mu$m. Shock age is calculated by dividing $\tau_{i}$ by $n_{e}$.}
\end{deluxetable}

\end{document}